\documentclass[conference,10pt]{IEEEtran}
\IEEEoverridecommandlockouts

\usepackage{amsmath,amsfonts}
\usepackage{algorithmic}
\usepackage{graphicx}
\usepackage{textcomp}
\usepackage{xcolor}

\usepackage{colortbl}
\usepackage{tikz}
\usepackage{xcolor}
\usepackage{cite}

\usepackage{pifont}
\usepackage{enumitem}

\def\BibTeX{{\rm B\kern-.05em{\sc i\kern-.025em b}\kern-.08em
    T\kern-.1667em\lower.7ex\hbox{E}\kern-.125emX}}

\newcommand{\shepherd}[1]{\textcolor{black}{#1}}

\newcommand{\reviseA}[1]{\textcolor{black}{#1}}

\newcommand{\reviseB}[1]{\textcolor{black}{#1}}

\newcommand{\reviseC}[1]{\textcolor{black}{#1}}

\newcommand{\reviseD}[1]{\textcolor{black}{#1}}

\newcommand{\reviseE}[1]{\textcolor{black}{#1}}

\newcommand{\reviseF}[1]{\textcolor{black}{#1}}

\newcommand*{\circled}[1]{\lower.7ex\hbox{\tikz\draw (0pt, 0pt) circle (.45em) node {\makebox[0em][c]{\small #1}};}}

\newcommand*{\circledb}[1]{\lower.7ex\hbox{\tikz\draw (0pt, 0pt) circle (.45em) node {\makebox[0em][c]{\footnotesize #1}};}}

\newcommand*\blackcircle[1]{\tikz[baseline=(char.base)]{
            \node[shape=circle,fill,inner sep=1pt,scale=0.8] (char) {\textcolor{white}{#1}};}}

\newcommand*\nofillcircle[1]{\tikz[baseline=(char.base)]{
            \node[shape=circle, draw=black, line width=0.5pt, inner sep=1pt, scale=0.8] (char) {\textcolor{black}{#1}};}}

\begin{document}

\pdfpagewidth=8.5in
\pdfpageheight=11in

\newcommand{\iscasubmissionnumber}{1232}

\pagenumbering{arabic}

\title{Accelerating MoE with Dynamic In-Switch Computing on Multi-GPUs}

\author{
    \IEEEauthorblockN{Qijun Zhang$^{1}$, Chen Zhang$^{2}$\textsuperscript{*}, Zhuoshan Zhou$^{2}$, Haibo Wang$^{3}$, Zhe Zhou$^{3}$, Zhipeng Tu$^{3}$, Guangyu Sun$^{4}$, \\ Zhiyao Xie$^{1}$, Yijia Diao$^{2}$, Zhigang Ji$^{2}$, Jingwen Leng$^{2,5}$, Guanghui He$^{2}$, Minyi Guo$^{2}$}
    \IEEEauthorblockA{Hong Kong University of Science and Technology$^1$, Shanghai Jiao Tong University$^2$, \\ Huawei Technologies Co. Ltd.$^3$, Peking University$^4$, Shanghai Qi Zhi Institute$^5$ \\ 
    qzhangcs@connect.ust.hk, \{chenzhang.sjtu, zs.zhou, diao\_yijia, zhigangji, guanghui.he\}@sjtu.edu.cn, eezhiyao@ust.hk, \\ \{wanghaibo33, zhouzhe22, tuzhipeng3\}@huawei.com, gsun@pku.edu.cn, \{leng-jw, guo-my\}@cs.sjtu.edu.cn}
}

\maketitle
\setcounter{page}{1}

\thispagestyle{plain}
\pagestyle{plain}

\begin{abstract}

Mixture-of-Experts (MoE) has been adopted by many leading large models to reduce computational requirements. However, frequent inter-GPU communication in MoE expert parallelism (EP) becomes a performance challenge. We observe substantial redundant inter-GPU data transfers in MoE that can be potentially addressed by in-switch computing. Unfortunately, the existing solution, NVLink SHARP (NVLS), can only support static collectives with regular patterns, incapable of dynamic communication with irregular patterns in MoE. 

To bridge the functionality gap, we propose DySHARP, an integral dynamic in-switch computing solution to accelerate MoE, encompassing both communication primitives and communication-aware scheduling: 1) Dynamic multimem addressing co-designs ISA, architecture, and runtime, as a dynamic extension to NVLS, reducing redundant traffic. However, the resulting traffic reduction is inherently asymmetric between two directions, preventing it from directly translating into speedup. 2) Token-centric kernel fusion deeply fuses the dispatch-computation-combine pipeline, resolving this asymmetry to translate traffic reduction into actual speedup. Compared with the state-of-the-art solution, DySHARP achieves up to 1.79× speedup. 

\end{abstract}

\begingroup\renewcommand\thefootnote{}
\footnotetext{* ~Chen Zhang is the corresponding author.}
\endgroup

\section{Introduction}

In recent years, the development of large language models has brought groundbreaking advances to many fields, including natural language processing~\cite{vaswani2017attention,llama4}, computer vision~\cite{dosovitskiy2020image,liu2021swin}, and reasoning~\cite{liu2024deepseek,yang2025qwen3}. To enhance model capabilities, the parameter count of models has been continuously scaling up~\cite{kaplan2020scaling}, leading to a significant surge in the computational demands required for model training. To reduce these computational requirements, many leading large models, including DeepSeek~\cite{liu2024deepseek}, GPT~\cite{gptoss,gpt5}, Llama~\cite{llama4}, Qwen~\cite{yang2025qwen3}, and Pangu~\cite{tang2025pangu}, have opted to adopt Mixture-of-Experts (MoE) architecture~\cite{lepikhin2020gshard} for model parameter scaling. Compared to the dense Transformer~\cite{vaswani2017attention}, MoE splits Feed-Forward Network (FFN) layer into multiple experts. Each token dynamically activates only a small subset of experts, enabling a substantial reduction in computational overhead. 

With the ever-increasing computational and memory demands, \emph{expert parallelism} (EP)~\cite{lepikhin2020gshard} that distributes experts across GPUs is proposed to train MoE on multi-GPU systems. However, since a token may activate experts on remote GPUs, EP requires frequent inter-GPU communication, including \emph{Dispatch} and \emph{Combine} communication operators. This frequent inter-GPU communication becomes a performance bottleneck in MoE execution~\cite{zhang2025comet,wang2025harnessing,hwang2023tutel,he2022fastermoe,rajbhandari2022deepspeed,shen2022se}, consuming 50-80\% execution time~\cite{zhang2025comet}.

\begin{figure}[!t]
\centering
\includegraphics[width=0.49\textwidth]{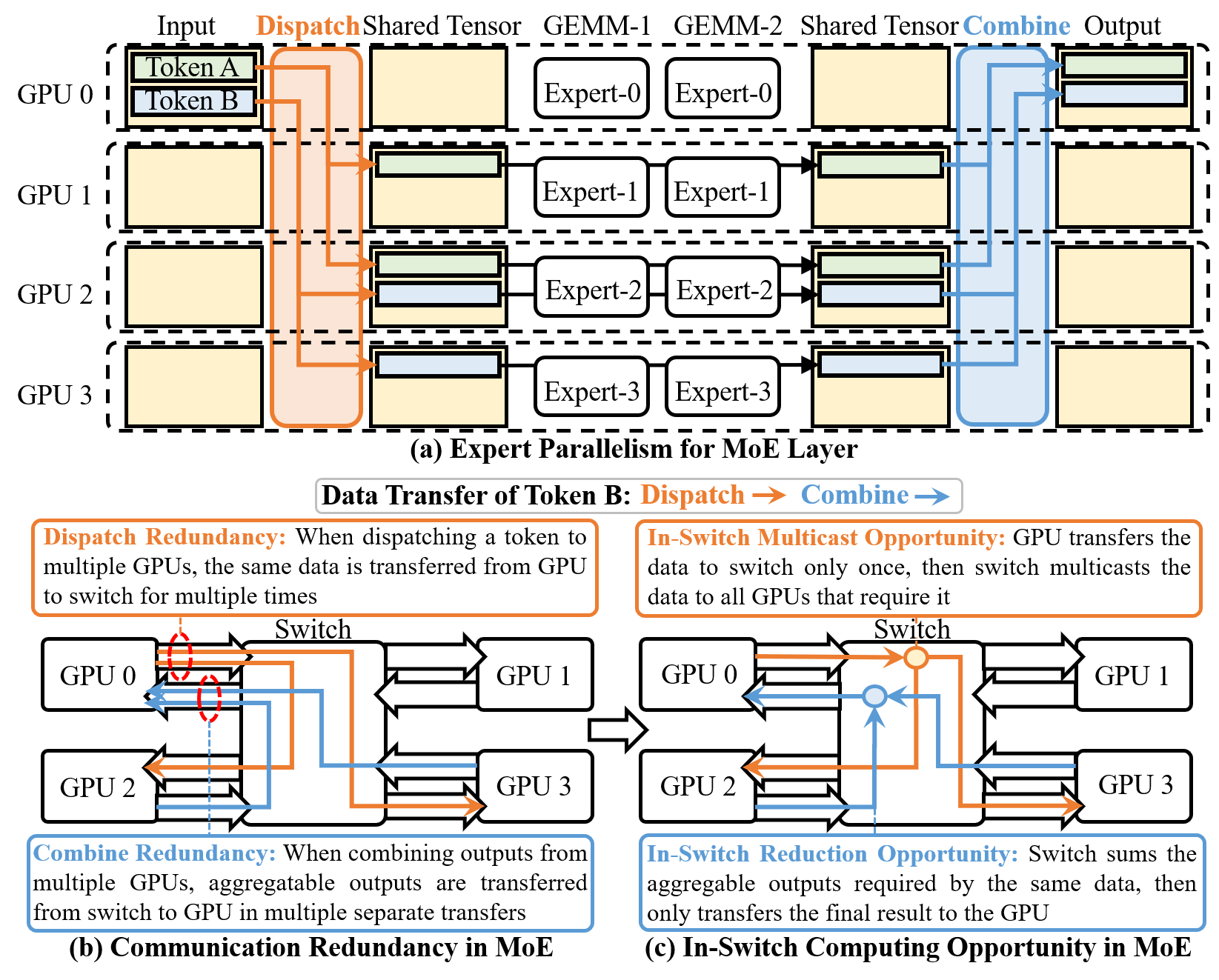}
\vspace{-.3in}
\caption{In-switch computing opportunity in MoE. MoE has significant redundant transfer that can be potentially addressed with in-switch computing.}
\label{opportunity}
\vspace{-.22in}
\end{figure}

We observe that both the Dispatch and Combine involve a fundamental communication inefficiency that no existing work tackles: redundant data movement across GPUs. As illustrated in Fig.~\ref{opportunity}(b), 1) When dispatching a token to multiple GPUs, the same data is transferred from GPU to switch multiple times. 2) When combining outputs from multiple GPUs, aggregatable outputs required by the same data are transferred from switch to GPU as multiple separate transfers. Profiling DeepSeek-v3 on a simulated GH200 NVL32~\cite{nvl32} shows a near 50\% communication redundancy out of total traffic. \looseness=-1

In-switch computing, which has been integrated in NVLink/ NVSwitch interconnection~\cite{nvl32,nvl72} through NVLink SHARP (NVLS)~\cite{klenk2020network}, \shepherd{can potentially address these redundancies through in-switch multicast and in-switch reduction}, as illustrated in Fig.~\ref{opportunity}(c). 1) With in-switch multicast, the GPU can only transfer the data to the switch once, then the switch multicasts the data to all GPUs, eliminating the dispatch redundancy. 2) With in-switch reduction, the switch can sum aggregable outputs and only transfer the final result to the GPU, eliminating the combine redundancy. 
However, despite the promising opportunity, the existing NVLS design is fundamentally static, restricted to the static collectives with regular communication patterns where the target sets are fixed and addresses are symmetric. Such a static NVLS design is thus incapable of accelerating dynamic operators with irregular communication patterns in MoE, with varying target sets and asymmetric addressing. A software-based workaround reinterpreting MoE communication as static collectives generates substantial useless traffic, which reaches 340\% in our profiling, negating the benefits of in-switch computing.

\shepherd{This functionality gap motivates us to propose a \emph{dynamic} in-switch computing solution to eliminate this communication redundancy in MoE. Achieving this requires not only communication primitives to reduce redundant traffic but also communication-aware scheduling to translate the reduction into actual speedup.} 
However, designing such a solution faces two challenges: \nofillcircle{1} The existing multi-GPU system design lacks top-down architectural support to enable the dynamic in-switch computing. \nofillcircle{2} Isolated dataflow schedule, where Dispatch and Combine are executed in isolation, incurs directional bandwidth imbalance, causing low overall bandwidth utilization. 
\shepherd{To address these challenges, DySHARP proposes an integral solution encompassing both communication primitives and communication-aware scheduling}: 
\blackcircle{1} Dynamic multimem addressing, a dynamic extension of NVLS's multimem addressing. Packet carries a single multimem address and a lightweight target list, with each GPU managing its memory locally. This supports irregularity of dynamic communication with high efficiency. 
\blackcircle{2} Token-centric kernel fusion to co-schedule operators. It utilizes token-level data dependency to pipeline the whole Dispatch-Computation-Combine chain. This token-paced pipeline merges complementary asymmetric communication patterns to improve bandwidth utilization. 
\shepherd{The two techniques work as an \emph{integral} solution, where neither alone is sufficient: through dynamic multimem addressing, DySHARP eliminates nearly half of the total traffic, but the resulting reduction is inherently asymmetric between two directions, preventing it from directly translating into speedup. Token-centric kernel fusion resolves this asymmetry, translating the traffic reduction into actual speedup and achieving complete redundancy elimination.} 
To the best of our knowledge, this is the first work to accelerate dynamic communication in MoE with in-switch computing. \looseness=-1

In summary, this paper makes the following contributions:
\begin{itemize}[leftmargin=0.4cm]
\item We conduct an in-depth analysis of the opportunity of leveraging in-switch computing to accelerate MoE's dynamic communication and the limitations of existing NVLS. 
\item We propose DySHARP, the first dynamic in-switch computing framework that introduces dynamic multimem addressing and token-centric kernel fusion to accelerate dynamic communication operators with fully exploited in-switch computing capabilities.
\item We evaluate DySHARP extensively under diverse workload configurations on a simulated GH200 NVL32 systems, demonstrating up to 1.79$\times$ speedup compared to the SOTA MoE acceleration solution.
\end{itemize}

\section{Background and Motivation}

\subsection{MoE Structure and Expert Parallelism}
\label{bgmoe}

MoE layer consists of multiple small FFNs each referred to as an expert, where each token only dynamically activates $topk$ experts selected by the gate network for computation, reducing computational cost. As illustrated in Fig.~\ref{opportunity}(a), each token is assigned to $topk$ experts, the assigned experts execute FFN computations (GEMM-1 and GEMM-2), and the $topk$ outputs are aggregated to produce the final result. \looseness=-1

To train MoE on multi-GPU systems, expert parallelism (EP) distributes experts across GPUs, introducing two inter-GPU communication operations as shown in Fig.~\ref{opportunity}(a): \emph{Dispatch} sends tokens to GPUs hosting their activated experts, and \emph{Combine} aggregates expert outputs back. This frequent communication becomes a performance bottleneck, accounting for 50-80\% of MoE layer execution~\cite{zhang2025comet}. Our experiment shows communication consumes 70.4\% of MoE layer execution in DeepSeek-V3 on simulated GH200 NVL32. Newer GPUs like NVIDIA Blackwell~\cite{gb200} and Rubin~\cite{rubin} are expected to see computational capacity outpace growth in communication, exacerbating this ratio. \looseness=-1

\subsection{Communication Redundancy in MoE}
\label{sec:redundancy}

\begin{figure}[!t]
\centering
\includegraphics[width=0.45\textwidth]{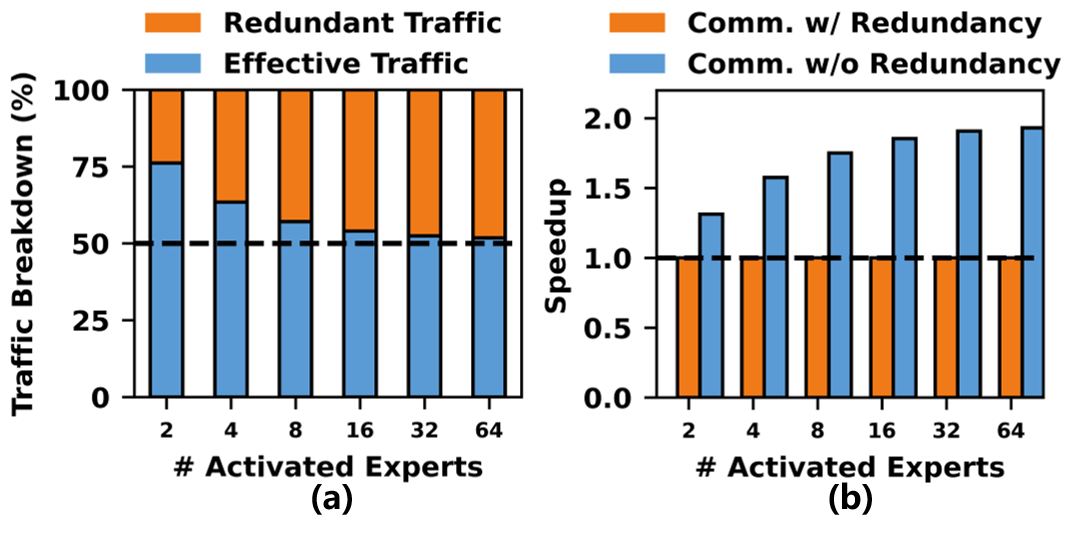}
\vspace{-.2in}
\caption{The quantification of (a) redundant data transfer and (b) acceleration opportunity of redundancy elimination of DeepSeek-V3 on a simulated GH200 NVL32-like 32-GPU system. It demonstrates significant potential for eliminating redundancy with in-switch computing.}
\label{redundancy}
\vspace{-.2in}
\end{figure}

A fundamental inefficiency of MoE communication lies in the large amount of redundant data movement across GPUs. In Dispatch, a single token often needs to reach multiple GPUs, where the same data is transmitted multiple times from source GPU to switch. For example, in Fig.~\ref{opportunity}(b), Token B on GPU0 must be sent to GPU 2 and 3, causing two identical but separate transfers over the GPU0-switch link. Similarly, in Combine, aggregatable expert outputs of a token from multiple GPUs are individually sent back, creating multiple separate transfers from switch to source GPU that the token originally resides. Intermediate outputs from both GPU 2 and 3 contribute to the output of Token B, causing two aggregatable but separate transfers over the switch-GPU0 link. 
Fig.~\ref{redundancy}(a) quantifies the redundant data transfer of DeepSeek-v3 with different numbers of activated experts on a 32-GPU system similar to the GH200 NVL32~\cite{nvl32}. It shows that there is significant redundant data transfer that accounts for nearly 50\% of the total traffic when the number of activated experts is over 8. \looseness=-1

Prior works~\cite{lepikhin2020gshard,nvidiamoe,shen2022se,deepep2025,hwang2023tutel,he2022fastermoe,wang2025harnessing,zhang2025comet} mitigate MoE's communication overhead through optimized libraries or computation-communication overlap, but none tackle the fundamental problem: redundant transfers of identical or aggregatable data.

\subsection{Opportunity and Limitation of In-Switch Computing}

\subsubsection{Redundancy Elimination with In-Switch Computing}
\label{sec:inswitchcomputing}

Theoretically, redundancies introduced in Sec.~\ref{sec:redundancy} can be eliminated through in-switch computing~\cite{graham2016scalable,sapio2021scaling,huang2025traci,klenk2020network,zhang2026towards}, which augments interconnect fabric with lightweight processing primitives. Since Hopper~\cite{h100}, NVIDIA has integrated in-switch computing in multi-GPUs interconnected via NVLink/NVSwitch with NVLink SHARP (NVLS)~\cite{klenk2020network}. \reviseC{NVLS introduces \emph{multimem} instructions to utilize in-switch computing for redundant transfer elimination: \texttt{multimem.st} accelerates AllGather through in-switch multicast, and \texttt{multimem.ld\_reduce} speeds up Reduce-Scatter via in-switch reduction.} 
This technique can also potentially address communication redundancy in MoE, as illustrated in Fig.~\ref{opportunity}(c): 

\begin{itemize}[leftmargin=0.4cm]
    \item \emph{In-switch multicast} eliminates the redundancy in the dispatch operator: Only a single copy of the data is transferred from the source GPU to the switch. The switch then multicasts it to all destination GPUs requiring that data. As shown in Fig.~\ref{opportunity}(c), GPU 0 sends the token to the switch only once. Switch multicasts the token to GPU 2 and 3. 
    \item \emph{In-switch reduction} eliminates the redundancy in the combine operator: The multiple intermediate outputs are accumulated within the switch. Only the final result is transferred from the switch back to the source GPU. The data from GPU 2 and 3 are aggregated within the switch, and only the final accumulated result is transferred to GPU 0. 
\end{itemize}
Therefore, with the capability of multicast and reduction inside the switch, the redundant data transfer can be eliminated. Fig.~\ref{redundancy}(b) quantifies the ideal acceleration opportunity with redundancy eliminated, indicating a significant ideal communication speedup near 2× with $\ge$8 activated experts.

\subsubsection{Limitation of Existing In-Switch Computing}
\label{sec:sharp}

\begin{figure}[!t]
\centering
\includegraphics[width=0.49\textwidth]{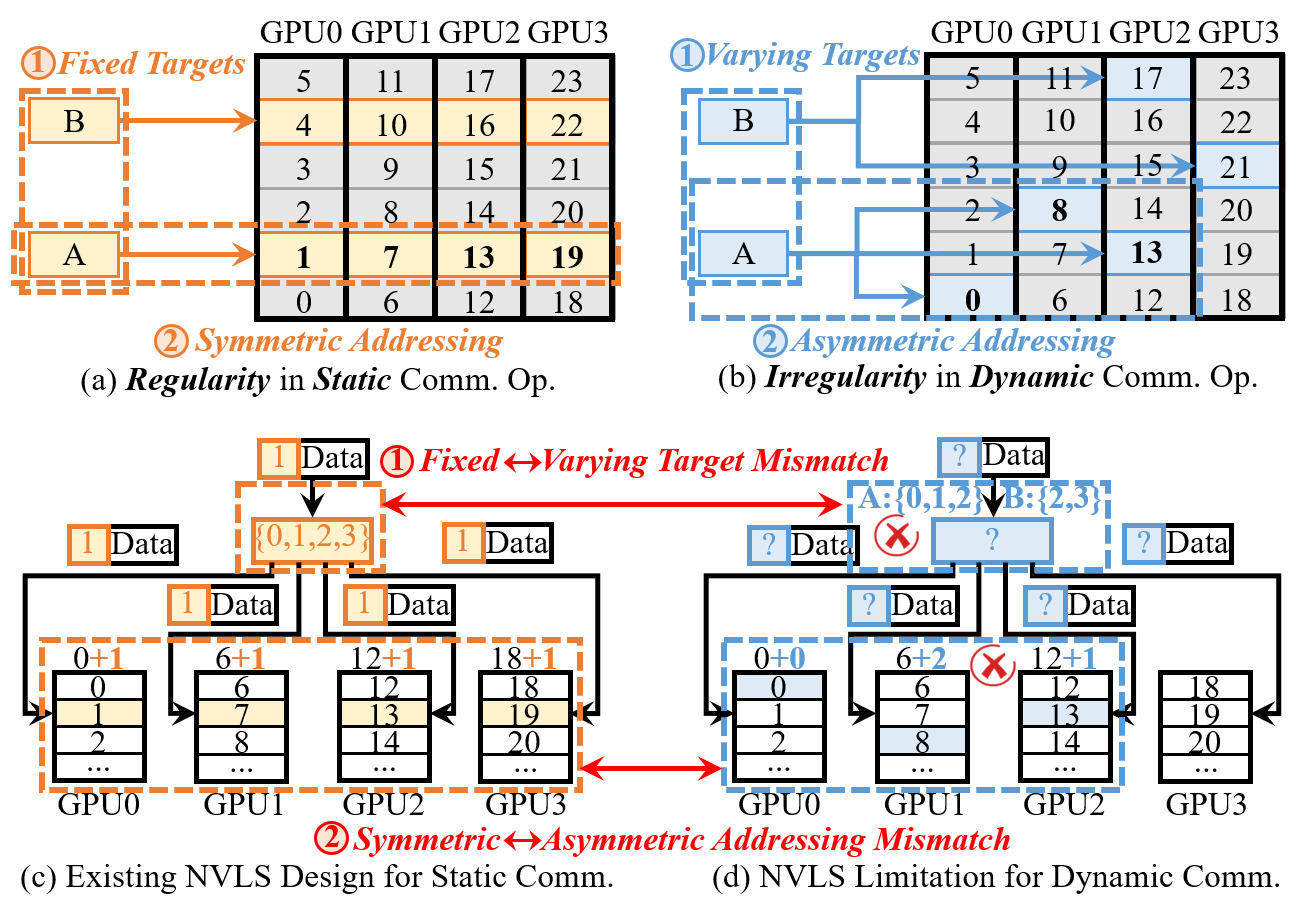}
\vspace{-.25in}
\caption{Communication pattern and NVLS applicability for static and dynamic communications. NVLS's customization for the regularity of static collectives leads to limitations for supporting dynamic communications with irregularity.}
\label{multimem}
\vspace{-.2in}
\end{figure}

While in-switch multicast and reduction are appealing to eliminate this communication redundancy, existing solution (NVLS) is fundamentally \textbf{static}, restricted to static collective operators like AllGather and Reduce-Scatter, and thus incapable of accelerating dynamic communication operators in MoE.

This limitation stems from NVLS's customization for the regularity inherent in static collective operations. As illustrated in Fig.~\ref{multimem}(a), this regularity manifests in two forms:
\begin{itemize}[leftmargin=0.4cm]
    \item Fixed target sets: All tokens of the operator always communicate with the same group of GPUs.
    \item Symmetric addressing: A token resides at identical memory offsets across GPUs.
\end{itemize}
This regularity enables NVLS to introduce multimem addressing, depicted in Fig.~\ref{multimem}(c): Packets carrying only one address and rely on preconfigured target set to determine destinations, minimizing header overhead for high bandwidth efficiency. 

However, such a NVLS cannot support the MoE's dynamic communication operators with the irregular pattern. The irregularity is shown in Fig.~\ref{multimem}(b): 
\begin{itemize}[leftmargin=0.4cm]
    \item Varying targets: each token may be routed to a different subset of experts, e.g., Token A $\rightarrow$ GPUs \{0, 1, 2\} while Token B $\rightarrow$ GPUs \{2, 3\}.
    \item Asymmetric addressing: tokens are independently allocated on each GPU in a dynamic approach, leading to divergent memory offsets, e.g., Token A are mapped to offsets \{0, 2, 1\} across GPUs \{0, 1, 2\}.
\end{itemize}
This communication pattern mismatch between dynamic communication in MoE and static collectives NVLS customized for makes NVLS unable to support MoE Dispatch and Combine: Preconfigured target set in switch cannot support varying targets, and carrying only one address in packet cannot determine the destinations under asymmetric addressing. \looseness=-1

A naïve workaround reinterprets MoE communication as static collectives: using AllGather to emulate Dispatch and Reduce-Scatter for Combine. However, this forces all GPUs to send/receive data irrespective of actual need, generating useless traffic. Profiling DeepSeek-V3 on a GH200 NVL32-like system reveals that this translation introduces 340\% useless traffic, negating the potential benefits of in-switch computing.

\subsection{Design Philosophy and Challenges}
\label{sec:challenge}

To bridge this functionality gap, we propose DySHARP. \shepherd{The core philosophy is to provide a \emph{dynamic} in-switch computing solution to eliminate communication redundancy}. However, designing such a system still faces two challenges:

\textbf{C1: Lack of Top-Down Architectural Support.} 
The existing multi-GPU system design lacks architectural support for such a dynamic in-switch computing. To address this problem, DySHARP introduces \textbf{dynamic multimem addressing}, a dynamic extension to the existing multimem addressing framework. Dynamic multimem addressing introduces top-down enhancement to the existing multi-GPU system, including packet format, ISA, microarchitecture of both GPU and switch, and CUDA runtime. This full-stack extension enables the functionality of dynamic in-switch computing. 
\shepherd{By eliminating redundant traffic via in-switch multicast and reduction, dynamic multimem addressing reduces nearly half of the total communication traffic.}

\begin{figure}[!t]
\centering
\includegraphics[width=0.49\textwidth]{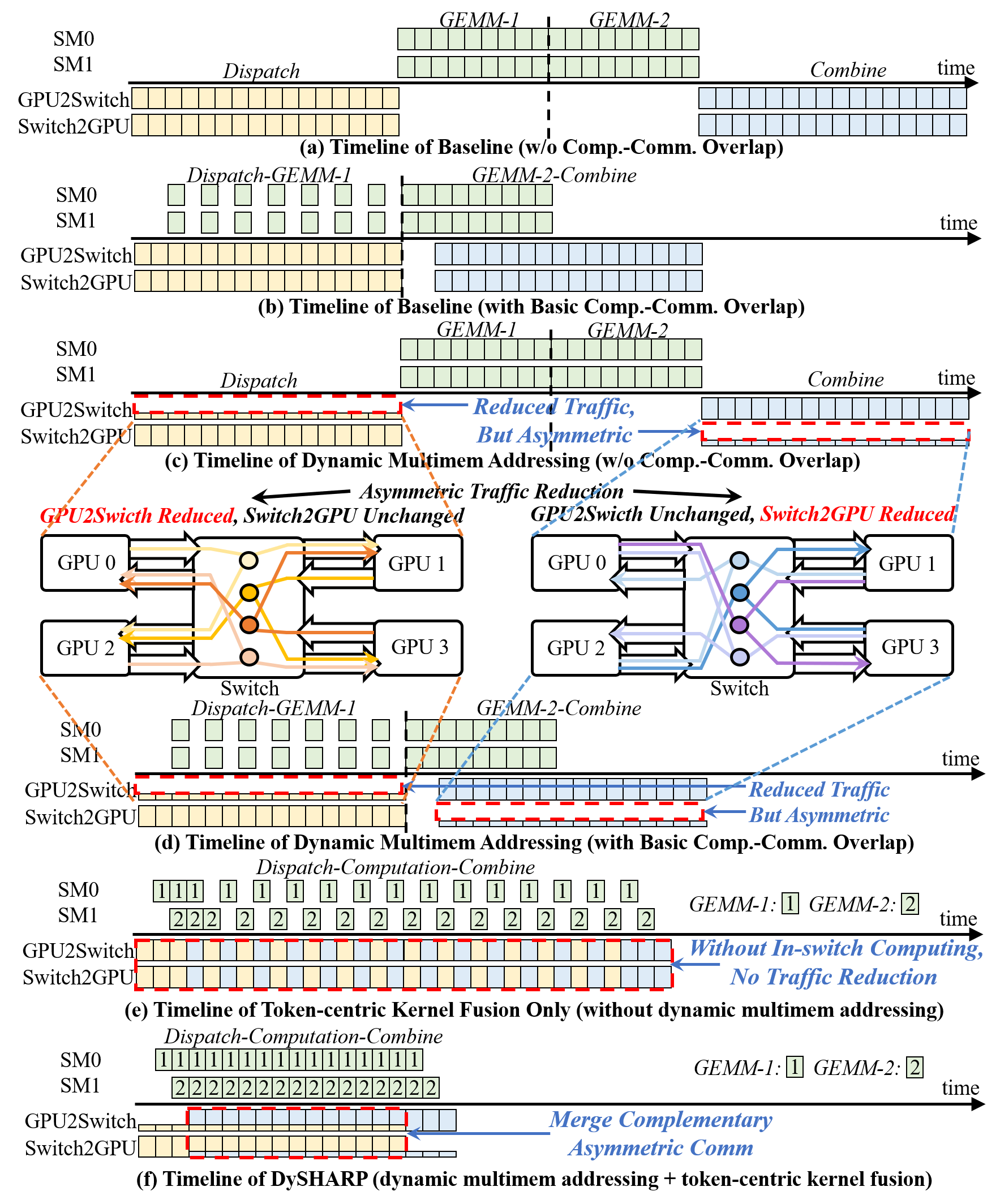}
\vspace{-.25in}
\caption{\shepherd{How the two techniques work as an integral solution. Dynamic multimem addressing reduces traffic but inherently introduces asymmetric reduction across directions. Token-centric kernel fusion resolves this asymmetry, translating traffic reduction into overall speedup. Neither alone is sufficient.}}
\label{motivation_overview}
\vspace{-.2in}
\end{figure}

\textbf{C2: Low Utilization of Isolated Dataflow Schedule.} 
Even with architectural support, executing dispatch and combine as isolated operators creates \emph{directional} bandwidth imbalance. In-switch multicast suppresses GPU$\rightarrow$switch traffic but leaves switch$\rightarrow$GPU heavy, and vice versa for reduction. The under-optimized direction dominates, resulting in low overall utilization. Overlap schemes~\cite{zhang2025comet,wang2025harnessing,hwang2023tutel,he2022fastermoe} can be composed with in-switch computing, but two communication kernels still remain isolated. 
DySHARP proposes \textbf{token-centric kernel fusion} to pipeline the whole Dispatch-Computation-Combine chain. This enables concurrent Dispatch and Combine, where complementary traffic patterns balance bidirectional bandwidth and significantly improve utilization. 
\shepherd{Fig.~\ref{motivation_overview} illustrates how the two techniques work as an integral solution. As shown in Fig.~\ref{motivation_overview}(b), in-switch computing inherently introduces traffic reduction that is asymmetric between two directions: in-switch multicast reduces GPU-to-switch traffic in Dispatch, while in-switch reduction reduces switch-to-GPU traffic in Combine. When Dispatch and Combine are executed in isolation, the unreduced direction becomes the bottleneck, preventing the traffic reduction from directly translating into speedup. Token-centric kernel fusion resolves this asymmetry, translating the traffic reduction into end-to-end speedup and achieving complete redundancy elimination, as shown in Fig.~\ref{motivation_overview}(d). Neither technique alone is sufficient: without traffic reduction, token-centric kernel fusion alone offers no improvement over existing techniques, as shown in Fig.~\ref{motivation_overview}(c). Detailed analysis is in Sec.~\ref{sec:overview2}, with quantitative validation in Sec.~\ref{sec:ablation}.}

\section{Dynamic Multimem Addressing}
\label{sec:dymultimemaddr}

\subsection{Key Idea of Dynamic Multimem Addressing}
\label{sec:overview1}

\begin{figure}[!t]
\centering
\includegraphics[width=0.47\textwidth]{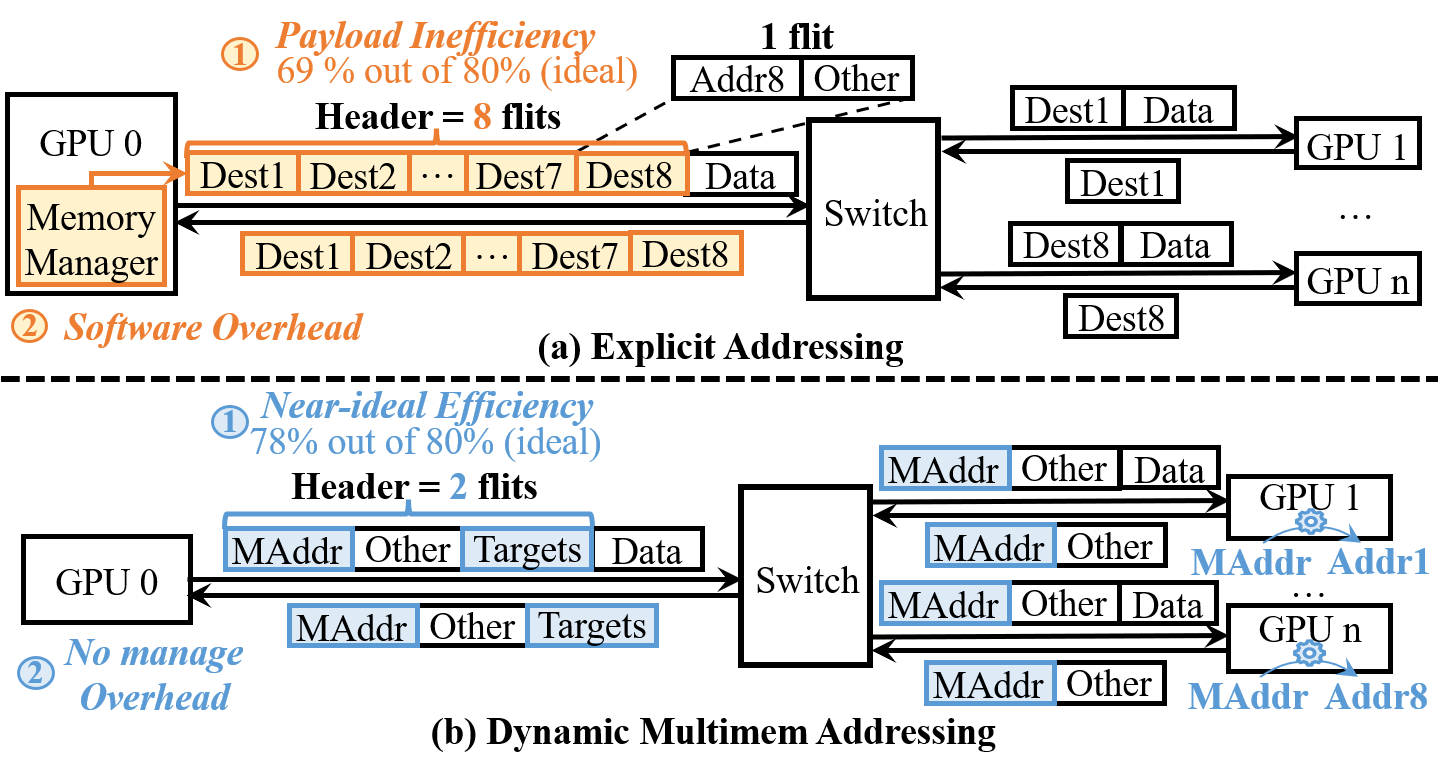}
\vspace{-.16in}
\caption{Two potential solutions for dynamic in-switch computing. (a) The straightforward solution is explicit addressing, but it has low payload efficiency and high software overhead. (b) Dynamic multimem addressing that we employ achieves near-ideal payload efficiency and no software overhead.}
\label{consideration}
\vspace{-.15in}
\end{figure}

To enable dynamic in-switch computing that supports the irregular dynamic communication in MoE, there are two potential solutions. The first solution is a straightforward approach that abandons multimem addressing and reverts to general shared-memory in-switch operations, which explicitly embeds all destination addresses, shown as explicit addressing in Fig.~\ref{consideration}(a). The second solution is to extend existing multimem addressing, still carrying only one address that represents multiple destinations of a request, which is our proposed dynamic multimem addressing, as shown in Fig.~\ref{consideration}(b). 

For explicit addressing, while it can support varying targets and asymmetric addressing, it is inefficient: 1) \emph{Payload inefficiency}: explicit destinations inflate packet headers and reduce payload efficiency; e.g., targeting eight GPUs requires eight destination flits in both request and response, dropping efficiency from an ideal 80\% to 69\%. 2) \emph{Software overhead}: sender must track remote memory states, e.g., per-expert token counters, and precompute destination addresses, causing extra synchronization (over 5\% performance loss~\cite{deepep2025}) and consuming 10-20\% of GPU compute resources~\cite{liu2024deepseek}. In comparison, dynamic multimem addressing is more promising. With only one address in the packet, it achieves high payload efficiency. Without considering detailed addressing in target GPUs, software overhead is also eliminated. Therefore, we employ dynamic multimem addressing in DySHARP. 

\begin{figure}[!t]
\centering
\includegraphics[width=0.43\textwidth]{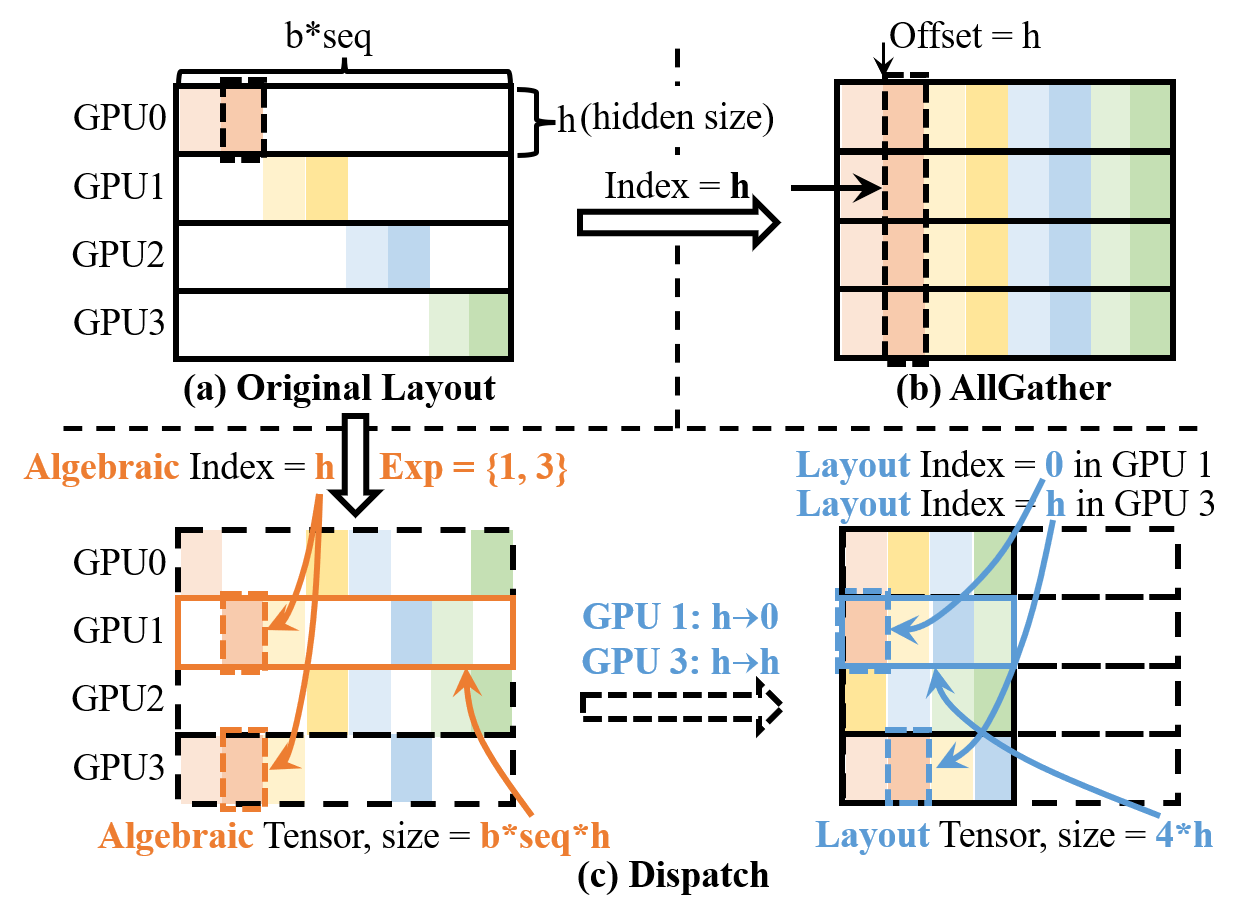}
\vspace{-.15in}
\caption{Comparison between Dispatch and AllGather. Dispatch/Combine are dynamic variants of AllGather/Reduce-Scatter, still with identical algebraic index across GPUs but per-GPU managed asymmetric memory layout.}
\label{overview1}
\vspace{-.15in}
\end{figure}

The core of designing dynamic multimem addressing is \emph{how to define the carried address that can represent the irregular multiple destinations of an operation.} We derive this address based on our understanding that Dispatch and Combine are the dynamic counterparts of AllGather and Reduce-Scatter. As shown in Fig.~\ref{overview1} (one expert per GPU), two properties follow:

\begin{enumerate}[leftmargin=0.4cm]

\item \textbf{Algebraic index is identical across GPUs.} In AllGather, each token is always broadcast to all GPUs and lands at the same index in the result tensor of each GPU. Algebraically, the only transformation from AllGather to Dispatch is that each token is sent only to a dynamically selected subset. Therefore, the \emph{algebraic index}, i.e., the index in the resulting \emph{algebraic tensor}, is still identical across GPUs.

\item \textbf{Memory layout is per-GPU managed and asymmetric.} Because only a subset of GPUs receives a given token, the algebraic tensor is fragmented and must be compacted into a dense \emph{layout tensor} to be stored in memory. 
This compaction is performed through stacking tokens within each GPU, so the resulting \emph{layout index}, i.e., the index in the layout tensor, is naturally asymmetric across GPUs.

\end{enumerate}
These properties give us the answer to question above, as illustrated in Fig.~\ref{consideration}(b): \emph{with a lightweight target expert list and an algebraic-layout mapping, the algebraic index can represent multiple destinations of an operation.} It drives two design points of our proposed dynamic multimem addressing:

\begin{itemize}[leftmargin=0.4cm]
    \item \textbf{Customized Packet}: A customized packet carries a single multimem address, whose offset is the algebraic index, and a target expert list. Packet format, ISA, and microarchitecture of GPU and switch are extended for support. 
    \item \textbf{Index Managing}: A hardware memory manager is proposed in Hub. This manager performs algebraic-layout index mapping to translate multimem address to virtual address, whose offset is layout index, for memory access. 
\end{itemize}
\reviseA{Our design introduces only minor, non-intrusive modifications to the existing hardware and software stack, building upon current datapaths without altering any original functionalities. This ensures low design complexity and minimal overhead, while preserving full support for other workloads.} 
Design details will be introduced in the following subsections.

\subsection{Packet Format Extension}
\label{sec:packet}

\begin{figure}[!t]
\centering
\includegraphics[width=0.48\textwidth]{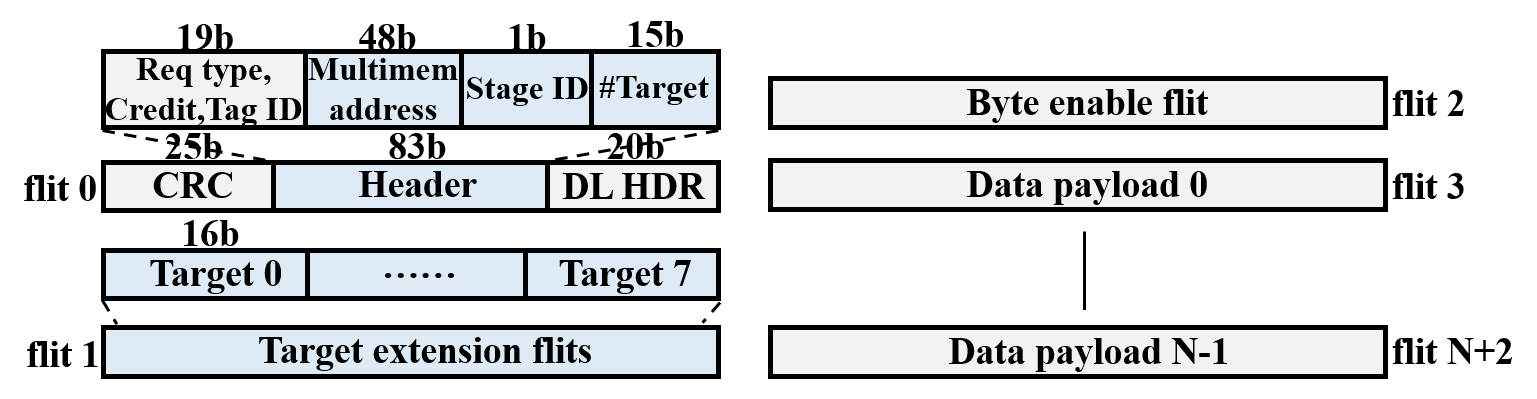}
\vspace{-.1in}
\caption{Our extended data link layer packet format for DySHARP based on the original NVLink packet format. Packet has only a single multimem address and an additional target list.}
\label{packet}
\vspace{-.05in}
\end{figure}

As introduced in Sec.~\ref{sec:overview1}, a customized packet format should be supported to carry a single multimem address and an additional target expert list. 
We extend NVLink data-link packet format to support dynamic multimem addressing framework, as shown in Fig.~\ref{packet}. In \emph{flit0}, we replace 64-bit address with: 1) 48-bit multimem address for algebraic index supporting 128TB address space, 2) 1-bit \emph{stage} (Dispatch/Combine), and 3) 15-bit \emph{target count}. Following \emph{flit0}, \emph{target extension} flits encode destination expert IDs (16\,bits each, eight per flit). Subsequent byte-enable and payload flits remain unchanged. Compared with explicit addressing that embeds full destination addresses, adopting such a multimem-style format preserves header compactness for near-ideal payload efficiency. 

\subsection{ISA Extension}
\label{sec:isa}

\begin{figure}[!t]
\centering
\includegraphics[width=0.44\textwidth]{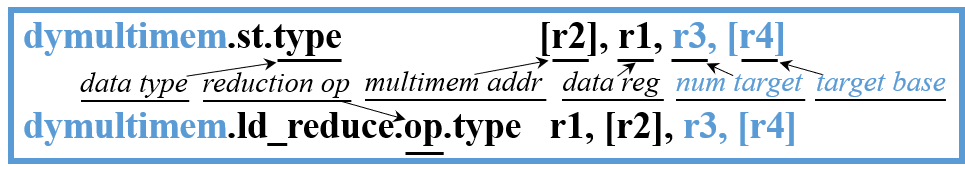}
\vspace{-.05in}
\caption{ISA extension for dynamic multimem addressing. We derive dymultimem instructions based on multimem instructions, with two additional registers required to specify the target list of the operation.}
\label{isa}
\vspace{-.2in}
\end{figure}

ISA should also be extended to provide a programming interface feasible for our packet extension. Because the request packet carries a single multimem address and a target expert list to support varying targets, ISA should provide this information, enabling source GPU to issue such a request packet.

\begin{figure*}[!t]
\centering
\includegraphics[width=0.95\textwidth]{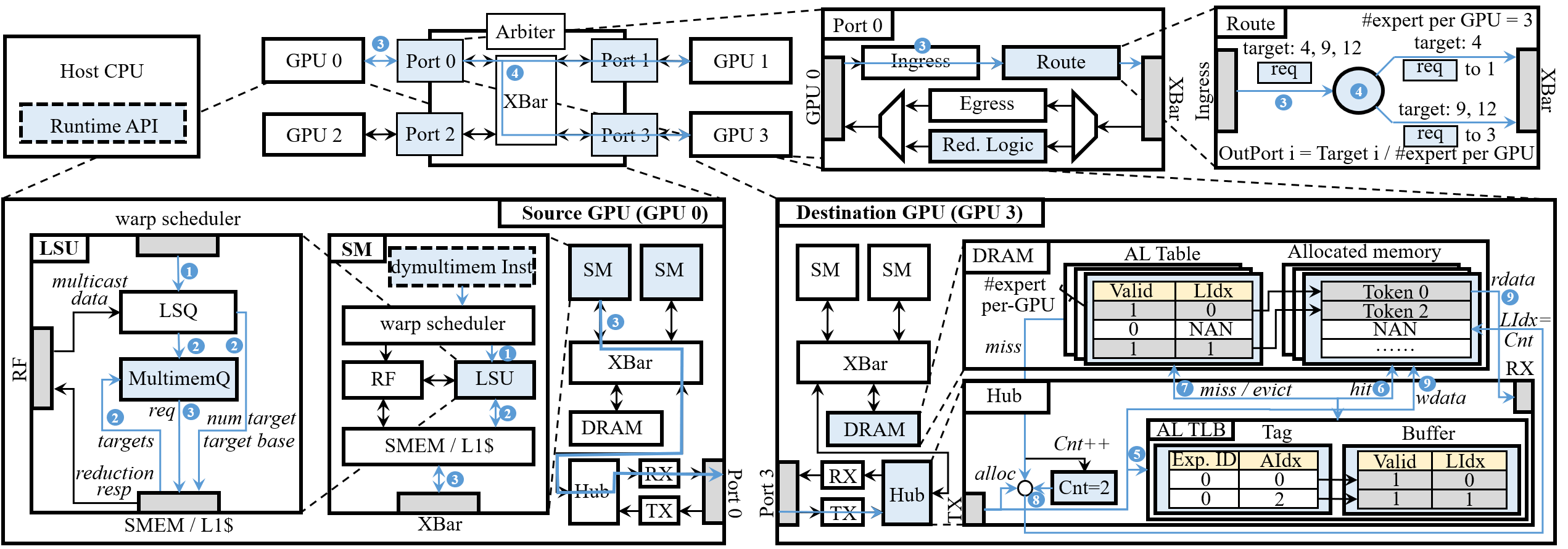}
\vspace{-.1in}
\caption{Detailed architectural design and workflow of the dynamic multimem addressing framework. 1) Source GPU includes a new LSU design in SM to fetch targets for dynamic multimem instructions. 2) Switch enhances forwarding and reduction, aware of the target list. 3) Destination GPU introduces a hardware memory manager in the Hub, performing multimem-virtual translation through mapping algebraic index to layout index. }
\label{architecture}
\vspace{-.2in}
\end{figure*}

Therefore, we introduce dymultimem instructions, an extension based on multimem instructions, as illustrated in Fig.~\ref{isa}. Extended instructions include \texttt{dymultimem.st} for \reviseF{multicast in Dispatch} and \texttt{dymultimem.ld\_reduce} for \reviseF{reduction in Combine}. Similar to original multimem instructions, dymultimem instructions use a register (\texttt{r2}) as multimem address, whose offset is the algebraic index, and another register (\texttt{r1}) to hold the data operand for \texttt{.st} or receive the reduced value for \texttt{.ld\_reduce}. \reviseB{Since NVLS's \texttt{multimem.ld\_reduce} does not support weighted reduction and adding weighting would incur high hardware complexity, our \texttt{dymultimem.ld\_reduce} retains reduction without weighting. Instead, we support weighted sum in Combine by applying weights in the epilogue of the preceding GEMM, before reduction.} 
\shepherd{Concretely, each expert scales its output $o_i$ by the gating weight $w_i$ in GEMM-2's epilogue, so the subsequent unweighted reduction $\sum_i(w_i \cdot o_i)$ yields the desired weighted sum.} \looseness=-1

As extension, to fetch target expert list, each instruction additionally specifies \texttt{r3} for the target count and \texttt{r4} for the base address of a contiguous target list. Targets can be fetched from global or shared memory, where list is usually loaded to shared memory in advance to reduce overhead.

\subsection{Microarchitecture Extension}
\label{sec:archdesign}

As illustrated in Fig.~\ref{architecture}, DySHARP introduces three components for the two design points, as introduced in Sec.~\ref{sec:overview1}. 

\begin{itemize}[leftmargin=0.4cm]
    \item For the customized packet, 1) source GPU\footnote{We call the GPU that a token resides on before Dispatch as source GPU of a token, and the GPU that a token is dispatched to as destination GPU.} includes new LSU design in SM to support target fetching for dynamic multimem instructions. 2) Switch enhances the forwarding and reduction, aware of the target list.
    \item For the index managing, 3) destination GPU introduces a hardware memory manager in the Hub. This manager performs multimem-virtual translation through mapping algebraic index to layout index locally within each GPU.
\end{itemize}

\subsubsection{Architectural Support in Source GPU}

We extend the SM’s LSU to fetch the target expert list and assemble dymultimem requests. Unlike regular memory instructions that read operands and immediately issue, a \texttt{dymultimem.*} first fetches its targets from memory. LSU uses target count and base pointer encoded in instruction to read the contiguous target list from shared or global memory. As shown in Fig.~\ref{architecture}, we introduce a small \emph{MultimemQ} separate from original LSQ to hold instructions that have issued their target fetching request. Once target list is ready, LSU issues the complete request packet to on-chip network and then to inter-GPU network. \looseness=-1

\subsubsection{Architectural Support in Switch}

Within the switch, the \emph{Route} module is augmented to forward dymultimem packets by their targets as shown in Fig.~\ref{architecture}. For each target $i$, the output port is computed as $\text{OutPort}_i = \text{Target}_i / \#\text{expert}\_\text{per}\_\text{GPU}$. Switch replicates the request per output port and trims each replica to include only the targets to the port. For \texttt{dymultimem.ld\_reduce} (Combine), the Reduction Logic records the number of targets associated with the request and decrements this counter as partial responses arrive; completion is detected when it reaches zero, which then returns the reduced result to the source GPU. This target-aware replication and completion tracking match the packet format in Sec.~\ref{sec:packet}.

\subsubsection{Architectural Support in Destination GPU}
\label{sec:archdestination}

At the destination GPU, a hardware memory manager performs algebraic-layout index mapping to translate the multimem address to a virtual address. This translation is performed before virtual-physical translation in GPU's Link MMU~\cite{ishii2022nvlink,klenk2020network}.

This manager performs translation with the token vector as its granularity. \reviseF{\emph{Dispatch:}} During Dispatch, the manager dynamically allocates the layout block to the algebraic block for arriving tokens sent by \texttt{dymultimem.st} request packets. The layout block is allocated in an accumulative approach, ensuring the fragmented algebraic tensor to be stored in a dense layout tensor. \reviseF{\emph{Combine:}} During Combine, with the algebraic-layout mapping built during Dispatch, the manager translates the algebraic index of \texttt{dymultimem.ld\_reduce} request packets to the layout index to load the data as responses.

\begin{figure}[!t]
\centering
\includegraphics[width=0.5\textwidth]{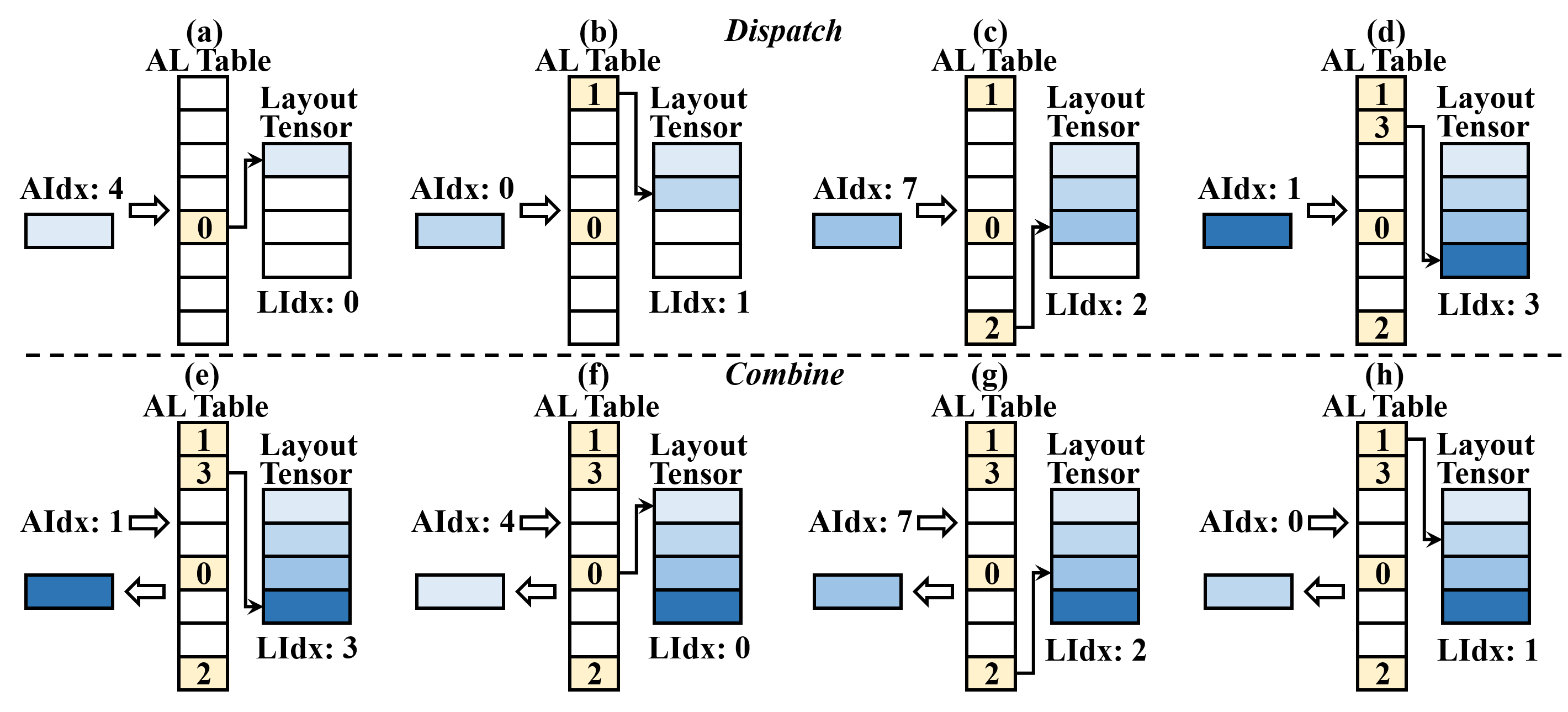}
\vspace{-.3in}
\caption{Illustration of hardware memory manager workflow for Dispatch and Combine, where AIdx is algebraic block index and LIdx is layout block index. Manager allocates a layout block to a new algebraic block during Dispatch, and then shares the same algebraic-layout mapping for Combine.}
\label{manager}
\vspace{-.2in}
\end{figure}

\medskip
\noindent\textbf{AL Management with AL Table.} 
The hardware memory manager performs algebraic-layout mapping (AL) management with an AL Table \reviseF{in GPU's DRAM.} AL Table is depicted in Fig.~\ref{architecture}, with each entry representing a mapping from an algebraic block to a layout block. The entry contains a \texttt{Valid} field to indicate whether a layout block has been allocated for this algebraic block, and a \texttt{LIdx} field to store the layout block index (LIdx). To find the layout block for an algebraic block, algebraic block index (AIdx) is used to index AL Table to get its LIdx. 
When multiple experts reside on a single GPU, the AL Table incorporates multiple independent sub-tables, with each sub-table separately managing the layout tensor of a distinct expert. 
\reviseF{Each AL Table entry is 4B (1b \texttt{Valid} + 31b \texttt{LIdx}), resulting in a table size of $4 \times \text{nToken}$ bytes. Even when processing 1M tokens, this consumption is only 4MB per layer, which is small compared to GPU DRAM (40s-100s GB).} \looseness=-1

Fig.~\ref{manager} illustrates how AL management is performed. \reviseF{\emph{Dispatch:}} Fig.~\ref{manager}(a)-(d) show the layout block allocation during Dispatch. When a new token, with an unseen AIdx of the \texttt{dymultimem.st} request packet, arrives, the manager allocates the next available layout block for it. The mapping is registered by writing the LIdx to the AIdx-th entry of the AL Table. A counter is adopted to track the next available layout block. 
\reviseF{\emph{Combine:}} Fig.~\ref{manager}(e)-(h) show the algebraic-layout mapping used by the Combine. Because the expert computation does not change the token order, the algebraic-layout mapping is unchanged before and after expert computation, leading the Combine to share the same AL Table as Dispatch. When a \texttt{dymultimem.ld\_reduce} request packet arrives, the manager looks up the AL Table to get LIdx to be accessed. 

\medskip
\noindent\textbf{MV Translation.} 
Hardware memory manager performs multimem-virtual address (MV) translation based on algebraic-layout mapping. For multimem address $MAddr$, AIdx is $(MAddr - MBase) / bsize$. After AL management resolves LIdx, virtual address $VAddr$ is $VBase +$ $LIdx * bsize + MAddr\,\%\,bsize$, where $MBase$ and $VBase$ are base addresses of multimem and virtual spaces. MV Translation is decoupled from AL management because Dispatch and Combine share the same algebraic-layout mapping but operate on different virtual address spaces. 

\medskip
\noindent\textbf{AL TLB.} 
Analogous to a conventional TLB, we introduce AL TLB to accelerate AL Table lookups. As shown in Fig.~\ref{architecture}, AL TLB uses the concatenation of Expert ID and AIdx as its tag. Each entry stores this tag along with its AL Table entry. The tag section employs Content-Addressable Memory (CAM) for fast lookup. The resulting index from tag matching accesses the buffer section implemented with SRAM. During access, 1) on a hit, the LIdx is directly got. 2) On a miss, the AL Table is accessed and the entry is brought into the AL TLB. In our software implementation for Dispatch and Combine, elements within the same token vector are typically accessed contiguously. This access pattern has strong temporal locality for AL TLB lookups, enabling a high AL TLB hit rate.

\subsubsection{Architectural Workflow}
\label{sec:archworkflow}

The workflow is depicted as Step \circled{1}--\circled{9} in Fig.~\ref{architecture}. On source GPU, \circled{1} a dymultimem instruction enters LSQ, then \circled{2} LSU fetches its target list from memory and \circled{3} issues complete dymultimem request through on-chip and inter-GPU networks to switch. \circled{4} Switch computes output ports per target and forwards replicated packets. At destination GPU, \circled{5} the request queries AL TLB: \circled{6} on a hit, MV translation directly yields the virtual address; \circled{7} on a miss, AL Table is accessed, with \circled{8} new layout blocks allocated on first touch during Dispatch. \circled{9} The translated virtual address is used for memory access: \texttt{dymultimem.st} writes data, and \texttt{dymultimem.ld\_reduce} reads and returns a response that is aggregated in the switch. 

\begin{figure}[!t]
\centering
\includegraphics[width=0.49\textwidth]{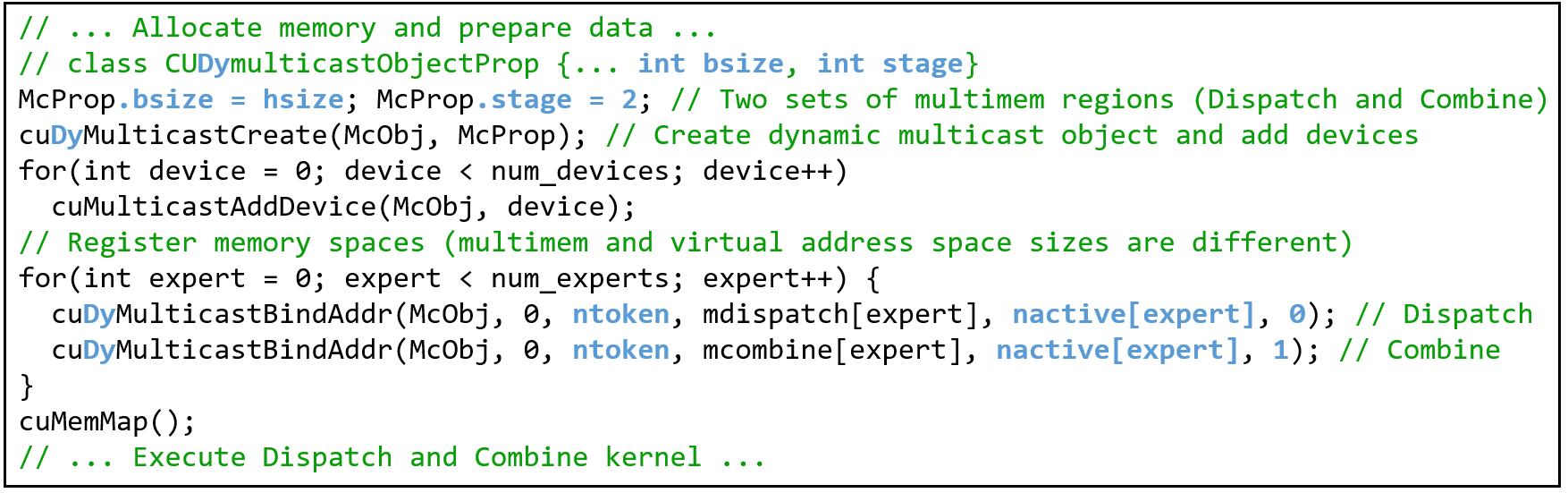}
\vspace{-.2in}
\caption{\shepherd{Code snippet with our extended CUDA Runtime API. We extend the existing CUDA Runtime API to support dynamic multimem addressing.}}
\label{runtime}
\vspace{-.17in}
\end{figure}

\subsection{Runtime Extension}
\label{sec:runtime}

We also extend the existing CUDA Runtime API for multicast object management to support the dynamic multimem addressing. Fig.~\ref{runtime} shows a CUDA code snippet utilizing the extended Runtime API. 
We introduce \texttt{CUDymulticastObjectProp} and its \texttt{cuDyMulti-} \texttt{castCreate} function. The extension specifies the block size \texttt{bsize} and \texttt{stage}, the number of multimem region sets sharing the same algebraic-layout index mapping, as the vector length \texttt{hsize} and 2 (Dispatch and Combine), respectively. 
We also extend the API with \texttt{cuDyMulticastBindAddr} to specify the sizes of the multimem and virtual address space, as \texttt{ntoken} and \texttt{nactive[expert]} in our example. 
\reviseF{The \texttt{nactive[expert]} represents the number of tokens to be dispatched to a given expert. This value is determined by token routing, which is generated by gating network \emph{before} Dispatch.} \looseness=-1

\section{Token-Centric Kernel Fusion}

Building on dynamic multimem addressing, we address low bandwidth utilization under isolated dataflow scheduling, as introduced in Sec.~\ref{sec:challenge}, with \emph{token-centric kernel fusion}, which co-schedules the full \emph{Dispatch-Computation-Combine} pipeline to co-execute Dispatch and Combine that have complementary communication patterns. Sec.~\ref{sec:overview2} outlines the key idea. Sec.~\ref{sec:tracker} presents a \emph{token tracker} that captures fine-grained, token-level dependencies across operators. Sec.~\ref{sec:scheduler} introduces a \emph{token-centric scheduler} that exploits these dependencies to pipeline operators, improving bandwidth utilization.

\subsection{Key Idea of Token-Centric Kernel Fusion}
\label{sec:overview2}

\begin{figure}[!t]
\centering
\includegraphics[width=0.49\textwidth]{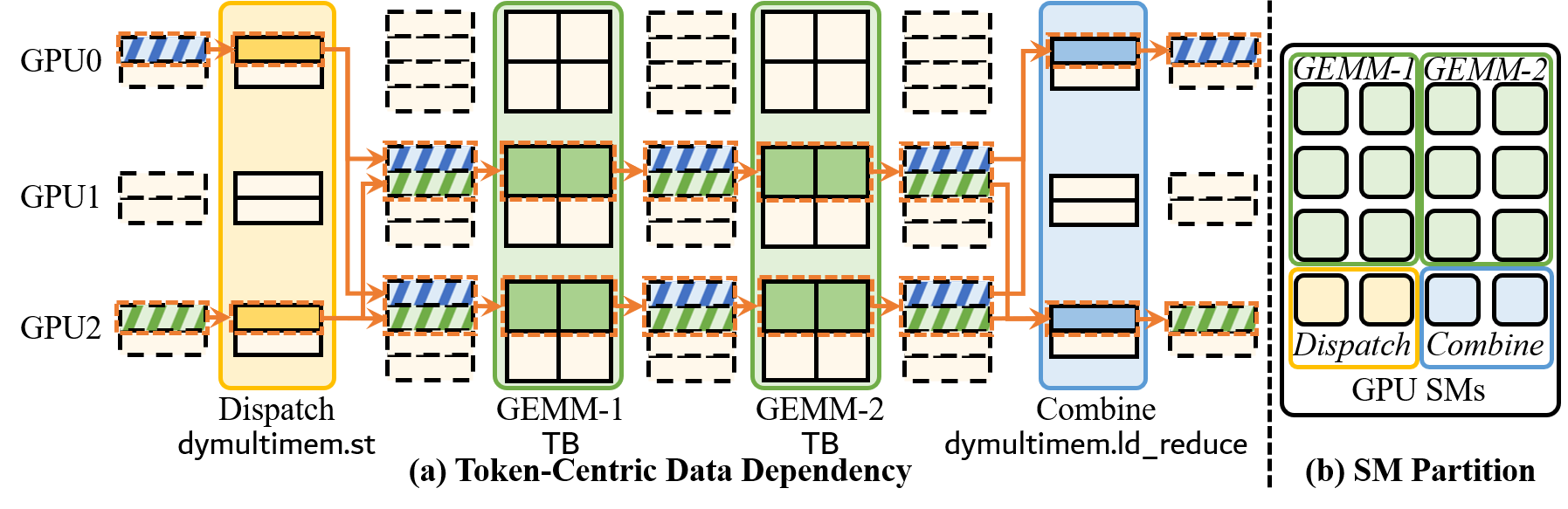}
\vspace{-.24in}
\caption{\shepherd{(a) Token-centric data dependency chain across Dispatch, GEMM-1, GEMM-2, and Combine. (b) SM partition for pipelined execution.}}
\label{dependency}
\vspace{-.15in}
\end{figure}

Token-centric kernel fusion treats the MoE layer as a \emph{token-paced pipeline} rather than four isolated operators. As illustrated in Fig.~\ref{dependency}(a), the insight is that readiness can be determined at \emph{token/tile} granularity, so operation can be performed as soon as their inputs for a given token (or a tile of $tsize$ tokens) become available, without waiting for operator-wide completion. Concretely, by explicitly tracking these token-level dependencies and scheduling at \emph{readiness boundaries}, Dispatch and Combine proceed \emph{concurrently}.

\shepherd{As previewed in Sec.~\ref{sec:challenge},} \shepherd{Fig.~\ref{motivation_overview}(a)–(f)} provides a detailed illustration of how kernel fusion \emph{translates the traffic reduction} of dynamic multimem addressing into \emph{speedup}. Without dynamic multimem addressing, the two baselines in \shepherd{Fig.~\ref{motivation_overview}(a)(b)}, i.e., DeepEP (baseline without overlap) and COMET (baseline with basic overlap), suffer from significant communication bottlenecks. Dynamic multimem addressing reduces GPU$\!\rightarrow$switch traffic for Dispatch and switch$\!\rightarrow$GPU traffic for Combine, yielding DySHARP-Basic (dynamic multimem addressing without overlap) and DySHARP-COMET (dynamic multimem addressing with basic overlap) in \shepherd{Fig.~\ref{motivation_overview}(c)(d)}. However, this traffic reduction does not directly lead to speedup because of the asymmetry between two directions in communication pattern. Hence we propose token-centric kernel fusion to translate traffic reduction into overall speedup. As depicted in \shepherd{Fig.~\ref{motivation_overview}(f)}, this is achieved by co-executing Dispatch and Combine concurrently, thereby merging complementary asymmetric communication patterns in \shepherd{Fig.~\ref{motivation_overview}(c)(d)}. This concurrent execution is fine-grained pipelining the \emph{whole} Dispatch-Computation-Combine flow. Importantly, token-centric kernel fusion \emph{alone} in \shepherd{Fig.~\ref{motivation_overview}(e)} does \emph{not} yield speedup over the SOTA baseline COMET, it must be \emph{integrated} with in-switch computing together to unlock the full performance potential. We provide detailed experimental analysis in Sec.~\ref{sec:ablation}. \looseness=-1

\subsection{Token Tracker}
\label{sec:tracker}

The token tracker is proposed to detect \emph{readiness boundaries}. It detects when a tile of tokens from Dispatch becomes consumable for its consumer GEMM TBs, and when a token with its $topk$ expert outputs are ready for its Combine.

\subsubsection{Token Tracker Design}

The tracker monitors the dependency chains as illustrated in Fig.~\ref{dependency}(a).

\begin{itemize}[leftmargin=0.4cm]
  \item \textbf{Dispatch$\Rightarrow$GEMM-1:}  
  When $tsize$ dispatched tokens for an expert have arrived through \texttt{dymultimem.st}, the corresponding row of GEMM-1 TBs is ready and can be issued immediately. Each expert locally counts arrived tokens. When counter reaches $tsize$, a row of GEMM-1 TBs is marked ready to issue. 
  \item \textbf{GEMM-1$\Rightarrow$GEMM-2:} A GEMM-2 TB row becomes ready when the corresponding GEMM-1 TB row completes. Tracker monitors TB completion of GEMM-1 and notifies the scheduler when a row of TB completes. 
  \item \textbf{GEMM-2$\Rightarrow$Combine:} 
  For each token, when all its $topk$ expert outputs are produced, Combine for this token can be executed via \texttt{dymultimem.ld\_reduce}. When a GEMM-2 TB row finishes its $tsize$ outputs, it notifies the source GPU of these tokens. Source GPU counts the number of notifications received for each token, and is ready when counter reaches $topk$.
\end{itemize}

\subsubsection{Token Tracker Architectural Support}
\label{sec:tracker}

\begin{figure}[!t]
\centering
\includegraphics[width=0.4\textwidth]{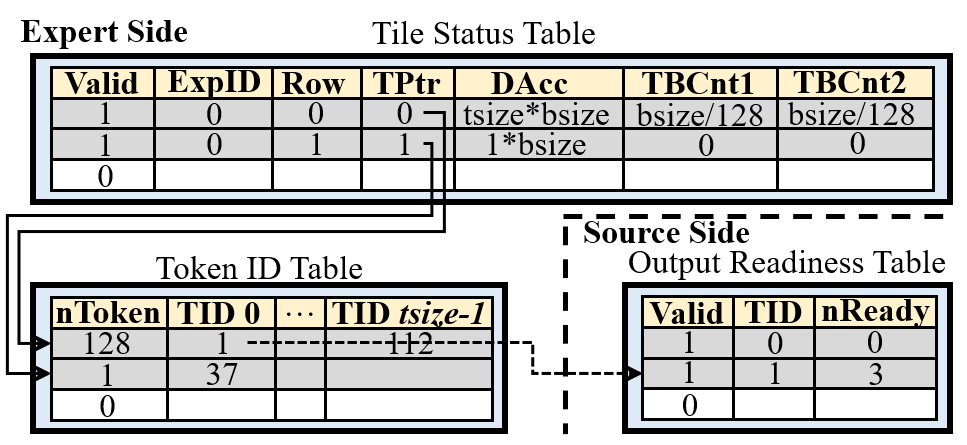}
\vspace{-.12in}
\caption{Architecture design of token tracker. Token tracker introduces Tile Status Table for \textbf{Dispatch$\Rightarrow$GEMM-1} and \textbf{GEMM-1$\Rightarrow$GEMM-2} readiness tracking, and cooperates with proposed Token ID Table and Output Readiness Table to track \textbf{GEMM-2$\Rightarrow$Combine} readiness.}
\label{tracker}
\vspace{-.15in}
\end{figure}

To implement the above readiness tracking, the tracker uses three lightweight tables, as shown in Fig.~\ref{tracker}.

\smallskip
\noindent\textbf{Tile Status (TS) Table} monitors the status of each $tsize$-token tile corresponding to a GEMM TB row. It tracks 1) the readiness of \textbf{Dispatch$\Rightarrow$GEMM-1}, 2) the readiness of \textbf{GEMM-1$\Rightarrow$GEMM-2}, and 3) the completion of GEMM-2 TB row to assist \textbf{GEMM-2$\Rightarrow$Combine} readiness tracking. 

Each entry TS Table includes a \texttt{Valid} field to indicate if the entry is valid, an \texttt{ExpID} field to identify the expert that the $tsize$ tokens belong to, and a \texttt{Row} field to record the row of TB these tokens correspond to. 
1) To track the readiness of \textbf{Dispatch$\Rightarrow$GEMM-1}, TS Table uses \texttt{DAcc} field to track the number of \texttt{dymultimem.st} access to the address region for these $tsize$ tokens.
Reaching $tsize*bsize$ indicates the arrival of dispatched $tsize$ tokens, marking the GEMM-1 TB row ready to issue. 
2) TS Table includes a \texttt{TBCnt1} field to track \textbf{GEMM-1$\Rightarrow$GEMM-2} readiness. This field tracks the number of completed GEMM-1 TB of this row. Completion of all TBs in this row indicates the readiness of the corresponding GEMM-2 TB row. 
3) Similar to \texttt{TBCnt1}, \texttt{TBCnt2} counts the number of completed GEMM-2 TB of this row. The completion of the row starts the notification to source GPUs for \textbf{GEMM-2$\Rightarrow$Combine} readiness tracking. The TS Table resides on-chip and can be offloaded to DRAM on overflow. 

\medskip
\noindent\textbf{Token ID (TID) Table} and \textbf{Output Readiness (OR) Table} are for \textbf{GEMM-2$\Rightarrow$Combine} readiness tracking. TID Table records tokens of each token tile, which are the tokens to be notified at the completion of GEMM-2 TB row. It records the number of tokens in this tile as \texttt{nToken} and each token ID as \texttt{TID}, which is registered when allocating the layout block to the algebraic block. Due to size and low access frequency, this table is placed in DRAM. 
OR Table receives this notification to track the readiness of each token for Combine. OR Table entry adopts a counter \texttt{nReady} to track the readiness of token \texttt{TID}. OR Table resides on-chip and can be offloaded to DRAM. 

When TS Table detects the completion of a GEMM-2 TB row, tracker notifies source GPUs of these tokens for their readiness. Tracker indexes TID Table with \texttt{TPtr} to collect completed token IDs, and then sends notifications to source GPUs of these tokens. When source GPU receives a notification, it increments \texttt{nReady} of the token entry. \texttt{nReady} reaching $topk$ indicates the readiness of this token for Combine. \looseness=-1

\reviseE{To guarantee visibility of produced data before accessing, all state updates in token tracker's tables are performed after written data is visible to all SMs, i.e., when the acknowledge is detected, indicating stored data has arrived at LLC/DRAM.}\looseness=-1

\subsection{Token-Centric Scheduler}
\label{sec:scheduler}

The scheduler realizes token-paced pipeline based on readiness detection of the tracker. It allows \texttt{Dispatch} and \texttt{Combine} to run concurrently, merging asymmetric traffic of in-switch computing. 
\reviseF{This scheduler is implemented in software through megakernel that employs persistent thread blocks (TBs) to bypass hardware TB scheduler~\cite{ma2020rammer}. Original TBs are represented as \emph{tasks}, and the action of \emph{issuing a TB to an SM} is emulated by a persistent TB fetching a task from the task list.} \looseness=-1

\subsubsection{SM Partitioning}
\label{sec:partition}
As shown in Fig.~\ref{dependency}(b), to achieve pipelining, SMs are partitioned into four groups dedicated to Dispatch, GEMM-1, GEMM-2, and Combine. A modified TB scheduler issues TBs of each kernel to its SM group. GEMM-1 and GEMM-2 can share SMs when one has no ready TB.

\subsubsection{Readiness-Gated Schedule}  
Consistent with Sec.~\ref{sec:overview2}, operation is gated by readiness besides resource availability. 
\reviseF{To check readiness for synchronization, the kernel polls the field indicating readiness in the token tracker's tables using a dedicated load instruction within a loop until they are ready:}

\begin{itemize}[leftmargin=0.4cm]
  \item \texttt{GEMM-1}/\texttt{GEMM-2}: a row of TB is issued only when the tracker marks the corresponding row \emph{ready} based on TS Table and the target SM group has capacity.
  \item \texttt{Combine}: communication kernels query token readiness, i.e., if the \texttt{nReady} of OR entry reaches $topk$, before issuing \texttt{dymultimem.ld\_reduce} for that token.
\end{itemize}

As shown in \shepherd{Fig.~\ref{motivation_overview}(f)}, because readiness is checked at token/tile granularity, MoE layer is executed as a token-paced pipeline. Dispatch, dominating GPU$\rightarrow$switch, and Combine, dominating switch$\rightarrow$GPU, naturally run in parallel, merging asymmetric pattern to improve bandwidth utilization.

\section{Experimental Methodology}

\subsection{Hardware Configuration}
\label{sec:setup}

In our experiment, we simulate the NVIDIA GH200 NVL32~\cite{nvl32}, a 32-GPU system interconnected via nine NVSwitch with a fully connected fat-tree topology. 
We integrate BookSim2~\cite{jiang2013detailed} and our customized Accel-Sim~\cite{khairy2020accel} to simulate our system \reviseC{in a cycle-accurate approach}, with each GPU configured based on the NVIDIA H200 specifications~\cite{h200}. \reviseC{For DeepSeek-V3, we extend the latest version of Accel-Sim, which supports basic Hopper features, to simulate high-performance FP8 kernels. To enable multi-GPU simulation, we support concurrent execution across GPUs connected through a switch-based network through BookSim2.} 

NVLink is modeled using real device parameters of NVLink 4.0~\cite{dgxh100}. The bidirectional bandwidth of NVLink is configured to 900 GB/s, and the latency of a single NVLink is configured to 250ns, where the round-trip latency is 1 µs. Flit size is set as 16B. For our modeled NVSwitch, each input port provides sixteen 256-depth virtual channels, with eight for requests and eight for responses. Port reduction buffer size is set to 64KB. 
For architectural support of dynamic multimem addressing, the MultimemQ in LSU consists of 32 entries, and AL TLB in Hub is configured to 512 entries. For token-centric kernel fusion, both TS Table and OR Table have 1024 entries. Our simulator is validated against DGX-H100, with average errors within 6\% for GEMM and DeepEP communication operators across diverse shapes/volumes.

\subsection{Benchmark}
\label{benchmark}

\begin{table}[!h]
      \centering
      \vspace{-.1in}
      \renewcommand{\arraystretch}{1.05}
      \resizebox{0.49\textwidth}{!}{
        \begin{tabular}{|c|c|c|c|c|c|c|}
\hline
\multicolumn{1}{|c|}{Name} & \multicolumn{1}{c|}{\begin{tabular}[c]{@{}c@{}}Hidden\\ Size\end{tabular}} & \multicolumn{1}{c|}{\begin{tabular}[c]{@{}c@{}}MoE Hidden\\ Size\end{tabular}} & \multicolumn{1}{c|}{\begin{tabular}[c]{@{}c@{}}Attention\\ Heads\end{tabular}} & \multicolumn{1}{c|}{\begin{tabular}[c]{@{}c@{}}Sequence\\ Length\end{tabular}} & \multicolumn{1}{c|}{\begin{tabular}[c]{@{}c@{}}Number of\\ Experts\end{tabular}} & \multicolumn{1}{c|}{\begin{tabular}[c]{@{}c@{}}$topk$\\ Candidates\end{tabular}} \\ \hline \hline
Small (S) & 2048 & 512 & 32 & 2048 & 64 & \{8, 16, 32\}\\ 
Medium (M) & 4096 & 1024 & 64 & 4096 & 128 & \{8, 16, 32\}\\ 
Large (L) & 7168 & 2048 & 128 & 8192 & 256 & \{8, 16, 32\}\\ 
\hline
\end{tabular}
        }
        \caption{Model Configurations adopted in evaluation.}
        \label{llmsetting}
        \vspace{-.2in}
\end{table}

DeepSeek-V3~\cite{liu2024deepseek} stands as one of the most competitive MoE-based LLMs. We therefore refer DeepSeek-V3 for our evaluation. Table~\ref{llmsetting} details the model configurations adopted in our evaluation. 
In addition to the official DeepSeek-V3 model configuration, denoted as \emph{Large} (L), we configure two additional model sizes: \emph{Small} (S) and \emph{Medium} (M). 
\reviseD{The number of activated experts, $topk$, is 8 in DeepSeek-V3. We also evaluate $topk=16/32$ to cover broader sparsity ranges, which may be potentially adopted in larger future models.} 
Our evaluation focuses on communication-heavy MoE training, with data parallelism for attention layers and expert parallelism for MoE layers \reviseD{within a NVL32 node. For end-to-end training, the 16-way pipeline parallelism is adopted across 16-NVL32 nodes~\cite{liu2024deepseek}.} 
\reviseB{Following ByteDance's observation for typical training jobs~\cite{zhang2025comet,cometcode}, we model the token distribution across experts as a normal distribution with a standard deviation ($std$) of 0.032.} \looseness=-1

\subsection{Baseline}
\label{sec:baseline}

DySHARP is evaluated against \reviseD{seven} baselines: 1) \textbf{DeepEP} ~\cite{deepep2025} is the state-of-the-art communication library for Dispatch and Combine, where no in-switch computing is utilized. 2) \textbf{NVLink SHARP (NVLS)}~\cite{klenk2020network} is the existing in-switch computing solution for static collective operations. Dispatch/Combine are replaced with AllGather/Reduce-Scatter as a workaround. 3) \textbf{FasterMoE}~\cite{he2022fastermoe} and 4) \textbf{Tutel}~\cite{hwang2023tutel} are coarse-grained computation-communication overlapping solutions for MoE. 5) \textbf{CCFuser}~\cite{wang2025harnessing} and 6) \textbf{COMET}~\cite{zhang2025comet} are fine-grained overlapping solutions, supporting Dispatch-GEMM and GEMM-Combine overlapping. \reviseD{7) \textbf{DualPipe}~\cite{liu2024deepseek} is an overlap strategy designed for cross-node pipeline.}

\section{Experimental Results}

\subsection{End-to-End and MoE-Layer Performance}
\label{sec:overallperf}

\subsubsection{End-to-End and MoE-Layer Speedup}
\label{sec:overallspeed}

\begin{figure}[!t]
\centering
\includegraphics[width=0.48\textwidth]{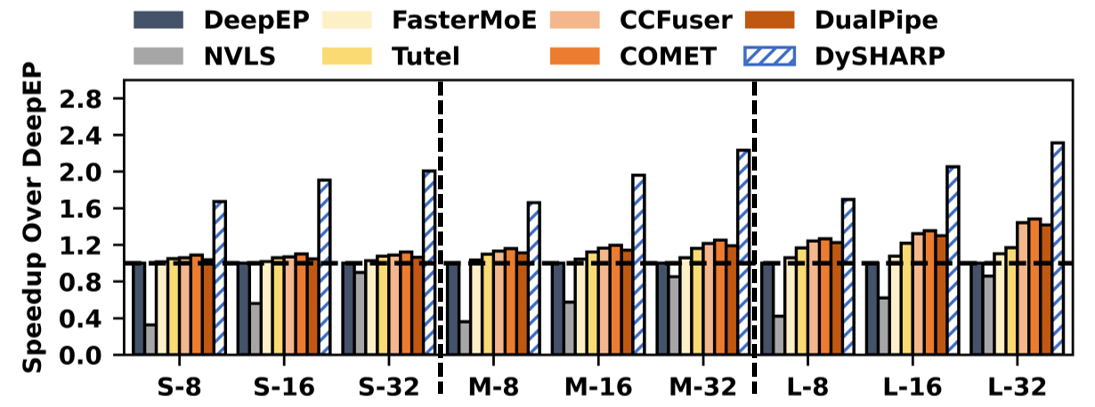}
\vspace{-.17in}
\caption{\reviseD{End-to-end model training speedup across different configurations.}}
\label{e2espeedup}
\vspace{-.05in}
\end{figure}

\begin{figure}[!t]
\centering
\includegraphics[width=0.48\textwidth]{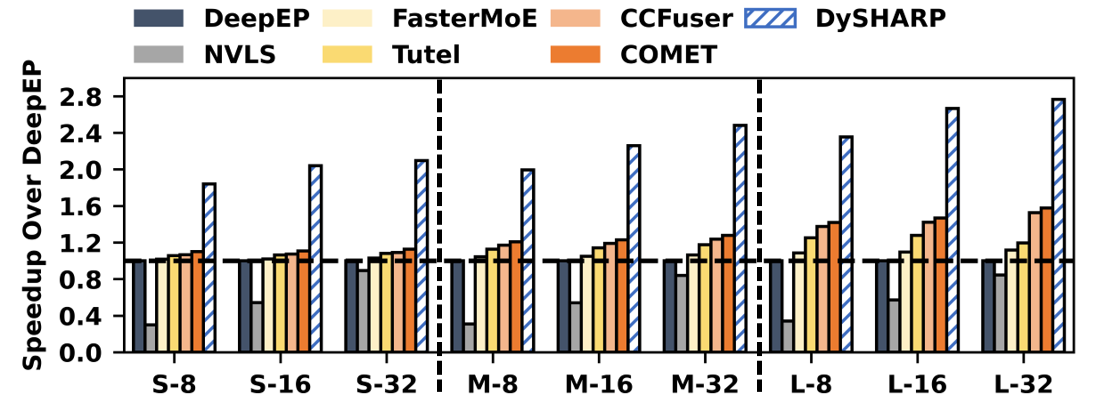}
\vspace{-.17in}
\caption{MoE layer speedup across different model configurations. \reviseD{(DualPipe is excluded because it is model-level cross-layer optimization.)}}
\label{opspeedup}
\vspace{-.15in}
\end{figure}

Fig.~\ref{e2espeedup} presents end-to-end model training speedup achieved by DySHARP compared with baselines across various model configurations and $topk$. \reviseE{This evaluation includes both attention and MoE layer, and covers both forward and backward propagation.} We denote configuration \texttt{Config} with $topk=k$ as \texttt{Config-k}. DySHARP achieves speedups of up to 2.31×, 5.12×, 2.11×, 1.98×, 1.85×, 1.79×, \reviseD{and 1.88×} over DeepEP, NVLS, FasterMoE, Tutel, CCFuser, COMET, \reviseD{and DualPipe} respectively, with geometric means of 1.93×, 3.38×, 1.84×, 1.72×, 1.63×, 1.59×, \reviseD{and 1.66×}. 
Fig.~\ref{opspeedup} further isolates performance comparison for communication-intensive MoE layer, encompassing Dispatch-Computation-Combine. \reviseD{DualPipe is excluded because it is model-level cross-layer optimization.} Compared with other six baselines, DySHARP achieves speedups of up to 2.77×, 6.93×, 2.48×, 2.32×, 2.01×, and 1.94×, respectively, with geometric means of 2.26×, 4.25×, 2.14×, 1.96×, 1.84×, and 1.78×. 
This demonstrates DySHARP's significant performance advantage, attributed to dynamic multimem addressing for redundant data transfer elimination and token-centric kernel fusion to merge asymmetric communication. 

\subsubsection{Discussions and Analysis}

\begin{figure}[!t]
\centering
\includegraphics[width=0.48\textwidth]{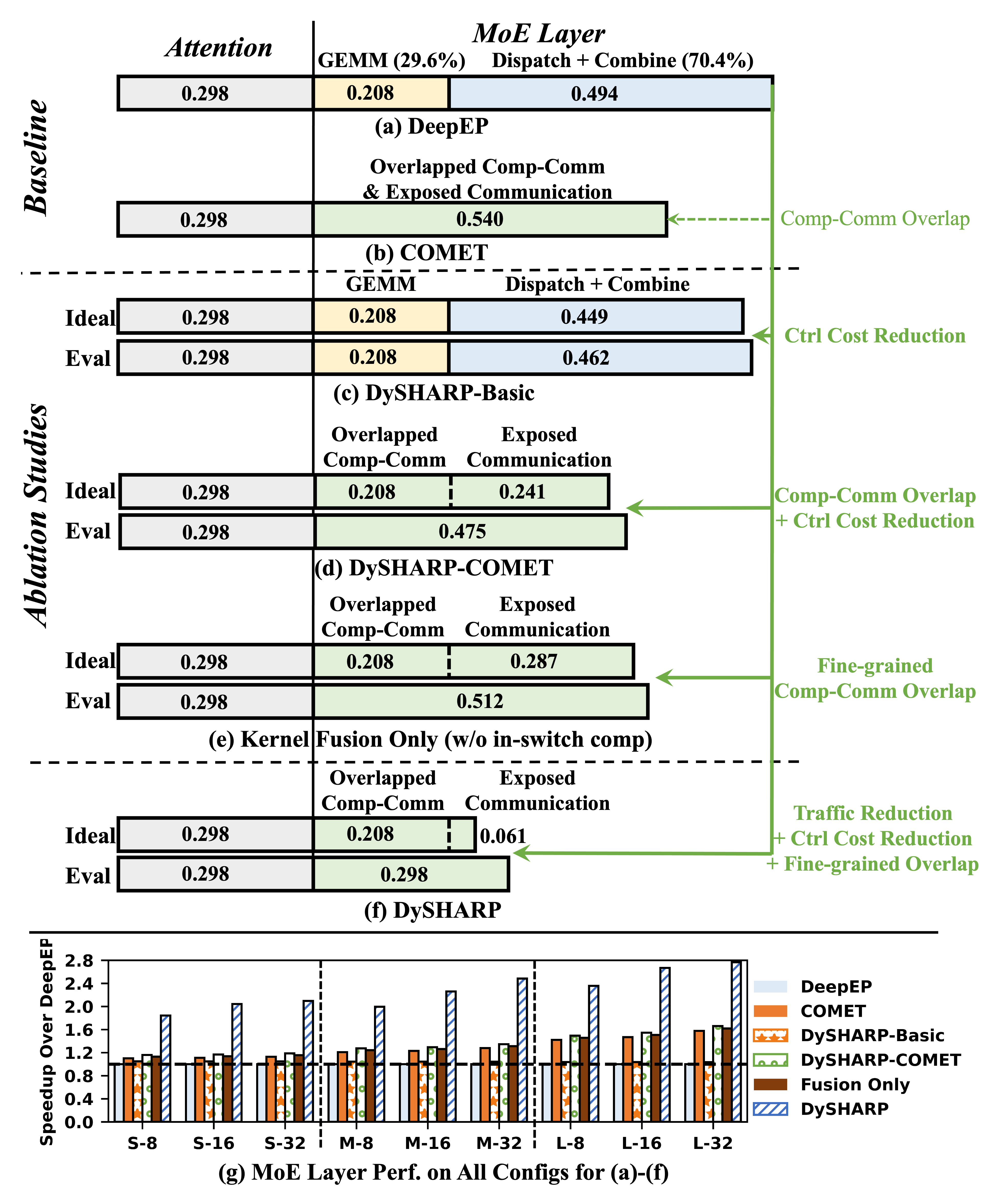}
\vspace{-.17in}
\caption{Quantitative time breakdown (normalized to DeepEP) and ablation studies on official DeepSeek-V3 configuration (L-8), validating \shepherd{Fig.~\ref{motivation_overview}(a)-(f)}.}
\label{breakdown}
\vspace{-.13in}
\end{figure}

DySHARP outperforms DeepEP, FasterMoE, Tutel, CCFuser, COMET, \reviseD{and DualPipe} primarily by eliminating redundant data transfers and reducing memory management overhead. Unlike these baselines with communication redundancy discussed in Sec.~\ref{sec:redundancy}, DySHARP leverages dynamic in-switch multicast and reduction to eliminate this redundancy, boosting performance. It also avoids software-controlled memory management and associated metadata transmission, e.g., token arrival counts, further reducing overhead. Fine-grained computation–communication overlap is also an advantage over baselines.

\begin{figure}[!t]
\centering
\includegraphics[width=0.49\textwidth]{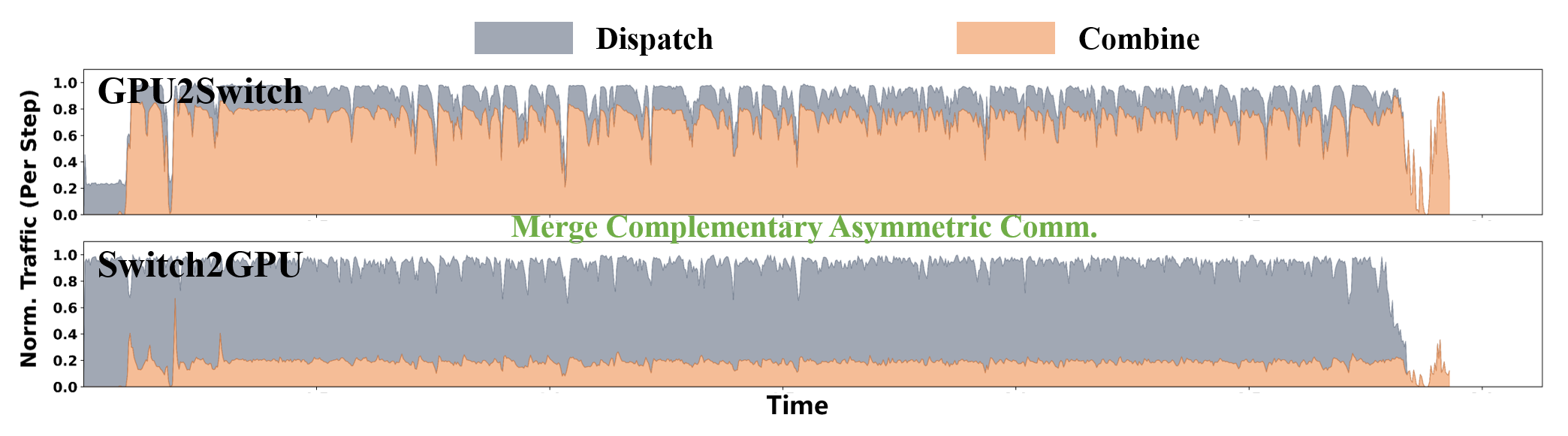}
\vspace{-.3in}
\caption{\reviseB{Illustration of merging complementary asymmetric communication.}}
\label{traffic2speedup}
\vspace{-.25in}
\end{figure}

DySHARP's performance advantage over the existing in-switch computing solution, NVLS, is from eliminating useless data transfers. 
This approach of replacing dynamic communication with static counterparts incurs large amounts of useless data transfer. In contrast, DySHARP can natively support dynamic communication without useless transfer. 

\subsubsection{\reviseB{Ablation Studies for Speedup Source Analysis}}
\label{sec:ablation}

We quantitatively validate the speedup sources analyzed in \shepherd{Fig.~\ref{motivation_overview}(a)–(f)}. In addition to DeepEP, COMET, and DySHARP, we further implement three variants for ablation study: 1) DySHARP-Basic (\shepherd{Fig.~\ref{motivation_overview}(c)}): dynamic multimem addressing without computation–communication overlap; 2) DySHARP-COMET (\shepherd{Fig.~\ref{motivation_overview}(d)}): DySHARP-Basic with COMET's overlap; and 3) kernel fusion only (\shepherd{Fig.~\ref{motivation_overview}(e)}): token-centric fusion without dynamic multimem addressing.

\reviseB{Fig.~\ref{breakdown}(a)–(f) shows the time breakdown on DeepSeek-V3 (Large-8). In Fig.~\ref{breakdown}(a)(b), DeepEP and COMET exhibit a severe communication bottleneck. With dynamic multimem addressing, DySHARP-Basic and DySHARP-COMET reduce traffic but do not directly lead to speedup as shown in Fig.~\ref{breakdown}(c)(d). This problem is due to asymmetric traffic reduction between the two directions. As an integral solution, DySHARP in Fig.~\ref{breakdown}(f) utilizes token-centric kernel fusion to merge complementary asymmetric communication by co-executing Dispatch and Combine concurrently, transforming traffic reduction enabled by dynamic multimem addressing into speedup. This merging of complementary communication can be observed in Fig.~\ref{traffic2speedup}. Moreover, kernel fusion \emph{alone} in Fig.~\ref{breakdown}(e) \emph{cannot} provide speedup over the SOTA baseline COMET, it must be integrated together with in-switch computing to unlock full potential. We also evaluate on all configurations, with results shown in Fig.~\ref{breakdown}(g).}

\subsection{Detailed Performance Analysis}

In this section, we individually analyze the effectiveness of dynamic multimem addressing and token-centric kernel fusion.

\subsubsection{Impact of Dynamic Multimem Addressing}
\label{impactdymultimem}

\begin{figure}[!t]
\centering
\includegraphics[width=0.48\textwidth]{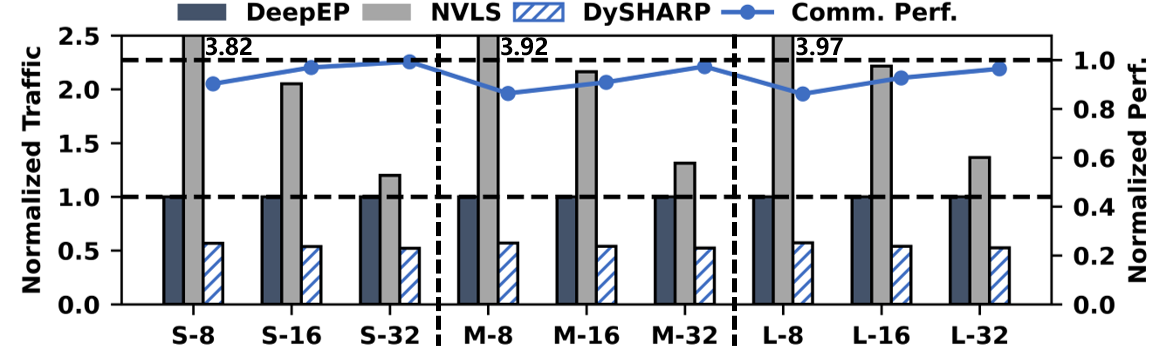}
\vspace{-.17in}
\caption{Traffic volume comparison and DySHARP communication capacity.}
\label{trafficreduction}
\vspace{-.15in}
\end{figure}

Dynamic multimem addressing reduces redundant data transfers in dynamic communication via in-switch computing, while avoiding useless data transfers present in existing in-switch computing solutions. Fig.~\ref{trafficreduction} compares the data transfer traffic of DeepEP, NVLS, and DySHARP, demonstrating DySHARP's effectiveness in significantly reducing data movement. 
Due to the substantial volume of useless data transfers, applying NVLS as a workaround results in increased data movement compared to DeepEP. DySHARP reduces traffic by nearly 50\% compared to DeepEP by eliminating redundant transfers. 
To inspect the communication capability of dynamic multimem addressing, we concurrently execute Dispatch and Combine operators without computation to measure the pure communication performance. Fig.~\ref{trafficreduction} demonstrates the pure communication performance normalized to the ideal calculated with traffic volume and bandwidth. DySHARP can, on average, achieve over 90\% performance of the ideal, indicating the high performance of such a dynamic multi-destination operation. 

\begin{figure}[!t]
\centering
\includegraphics[width=0.48\textwidth]{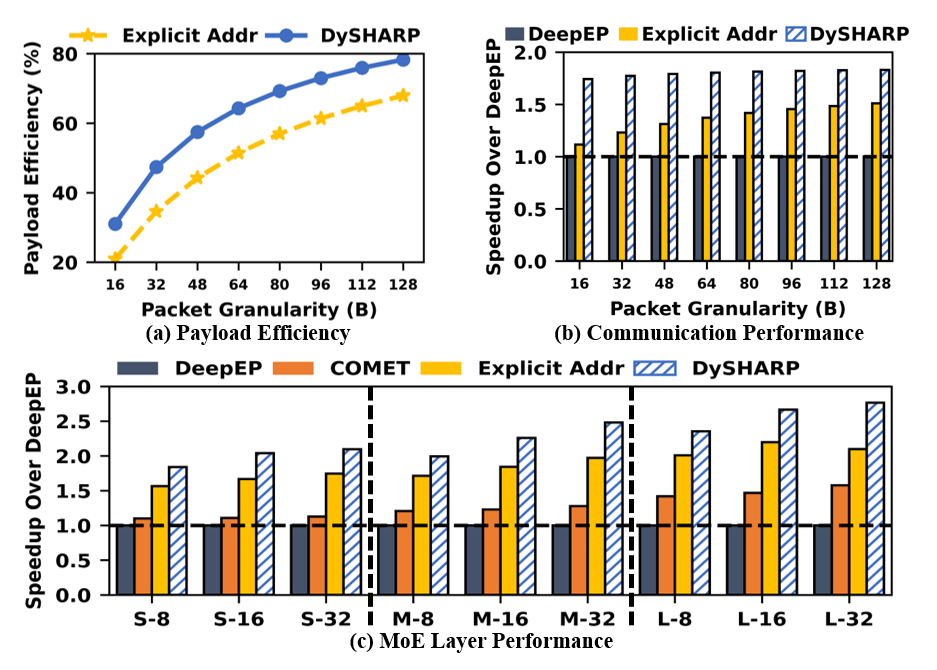}
\vspace{-.2in}
\caption{Comparison between DySHARP and explicit addressing on (a) payload efficiency, (b) communication, \reviseB{and (c) MoE layer performance.}}
\label{packetgran}
\vspace{-.15in}
\end{figure}

We analyze advantage of dynamic multimem addressing compared to straightforward explicit addressing that explicitly encodes all destinations within request packet. Fig.~\ref{packetgran}(a) compares payload efficiency (the proportion of data flits to total transmitted flits) under different data transfer granularities when targeting 8 destinations. Results demonstrate DySHARP's consistently higher payload efficiency than explicit addressing. Fig.~\ref{packetgran}(b) further compares performance of pure communication operators, highlighting the gain from high payload efficiency. 
\reviseB{We also adapt token-centric kernel fusion for explicit addressing and evaluate MoE layer performance. Fig.~\ref{packetgran}(c) shows results, validating DySHARP's advantage.} 
These results verify discussion in Sec.~\ref{sec:overview1}. 

\begin{figure}[!t]
\centering
\includegraphics[width=0.47\textwidth]{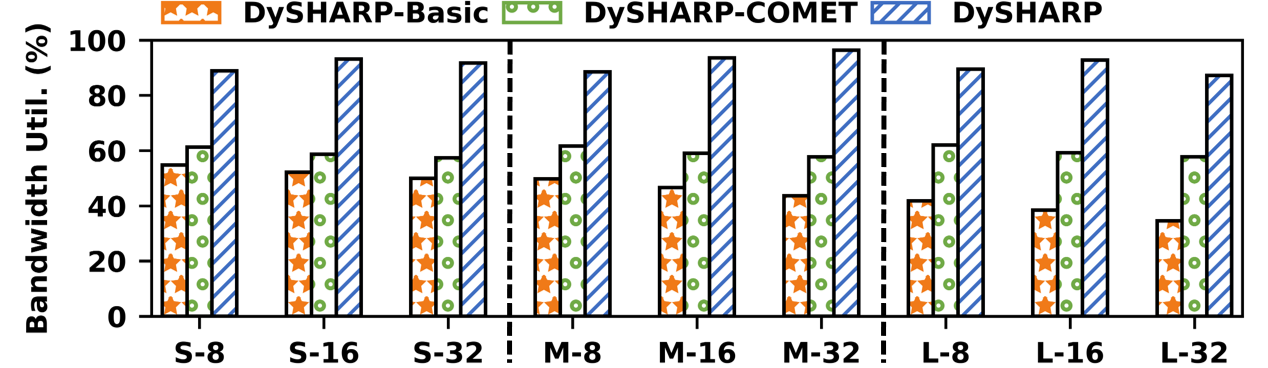}
\vspace{-.12in}
\caption{Bandwidth utilization comparison. Token-centric kernel fusion improves utilization over non-overlap and the SOTA overlap solutions.}
\label{utilization}
\vspace{-.15in}
\end{figure}

\subsubsection{Impact of Token-Centric Kernel Fusion}
\label{sec:impactfusion}

Token-centric kernel fusion improves overall bandwidth utilization by enabling token-paced pipeline of the Dispatch-Computation-Combine workflow. Fig.~\ref{utilization} compares the bandwidth utilization of full DySHARP against DySHARP-Basic and DySHARP-COMET. 
Without token-centric kernel fusion, DySHARP-Basic exhibits low bandwidth utilization due to 
non-overlapped computation-communication and asymmetric communication. While DySHARP-COMET achieves overlap, isolated Dispatch and Combine are still asymmetric. \reviseB{DySHARP with token-centric kernel fusion merges asymmetric Dispatch and Combine, transforming traffic reduction into speedup.} \looseness=-1

\subsection{Sensitivity Analysis}

\subsubsection{Performance of Different Numbers of GPUs}
\label{sec:ngpu}

\begin{figure}[!t]
\centering
\includegraphics[width=0.48\textwidth]{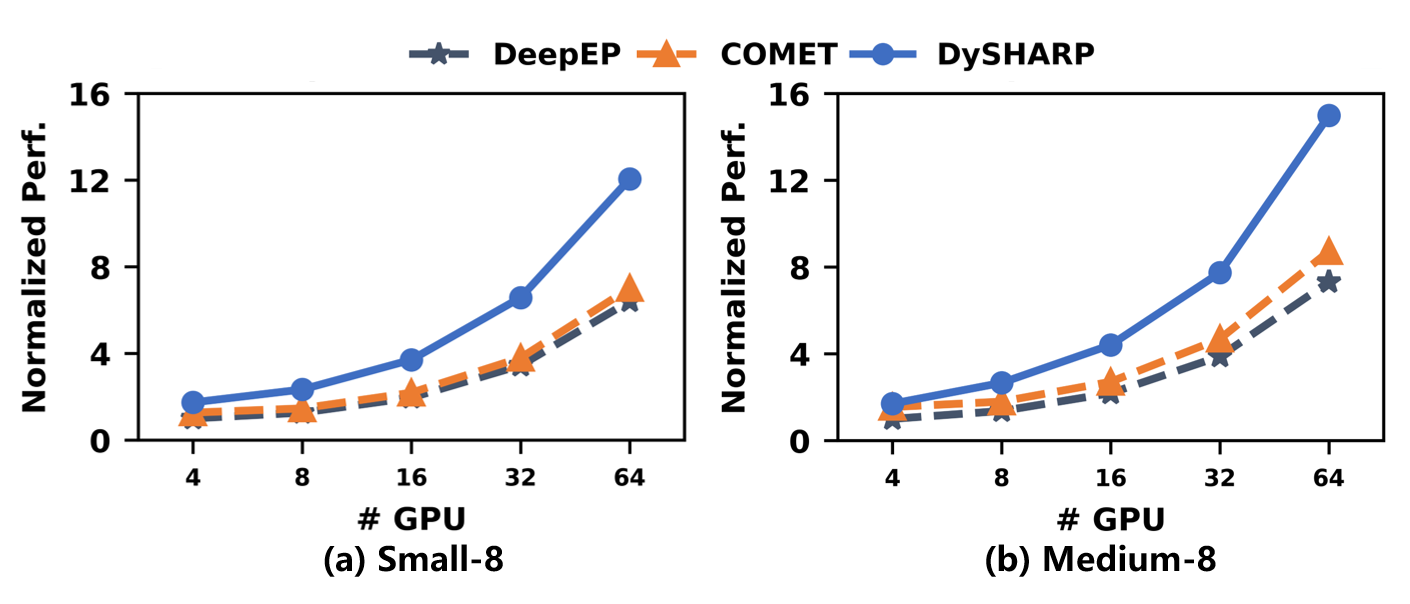}
\vspace{-.2in}
\caption{Performance sensitivity to the number of GPUs.}
\label{nGPU}
\vspace{-.17in}
\end{figure}

We evaluate the performance of DySHARP against baselines across different system scales. Using Small-8 and Medium-8 model configurations, we compare DySHARP with DeepEP and COMET across GPU counts of 4-64. Fig.~\ref{nGPU} presents our evaluation on MoE layer. \reviseD{We simulate the 64-GPU node as an extension of NVL32, where the only difference is the interconnect. We simulate such a system with doubled number of NVSwitch (18 NVSwitch). Each NVSwitch has 64 ports, and each port is connected to a GPU, providing full bandwidth for the GH200 chip that has 18 ports.} 
The results show that as the number of GPUs increases, DySHARP consistently outperforms both DeepEP and COMET, with the gap progressively widening. This highlights DySHARP's strong scalability and demonstrates its potential for future larger-scale SuperPODs.

\subsubsection{Performance of Different Sequence Lengths}

\begin{figure}[!t]
\centering
\includegraphics[width=0.48\textwidth]{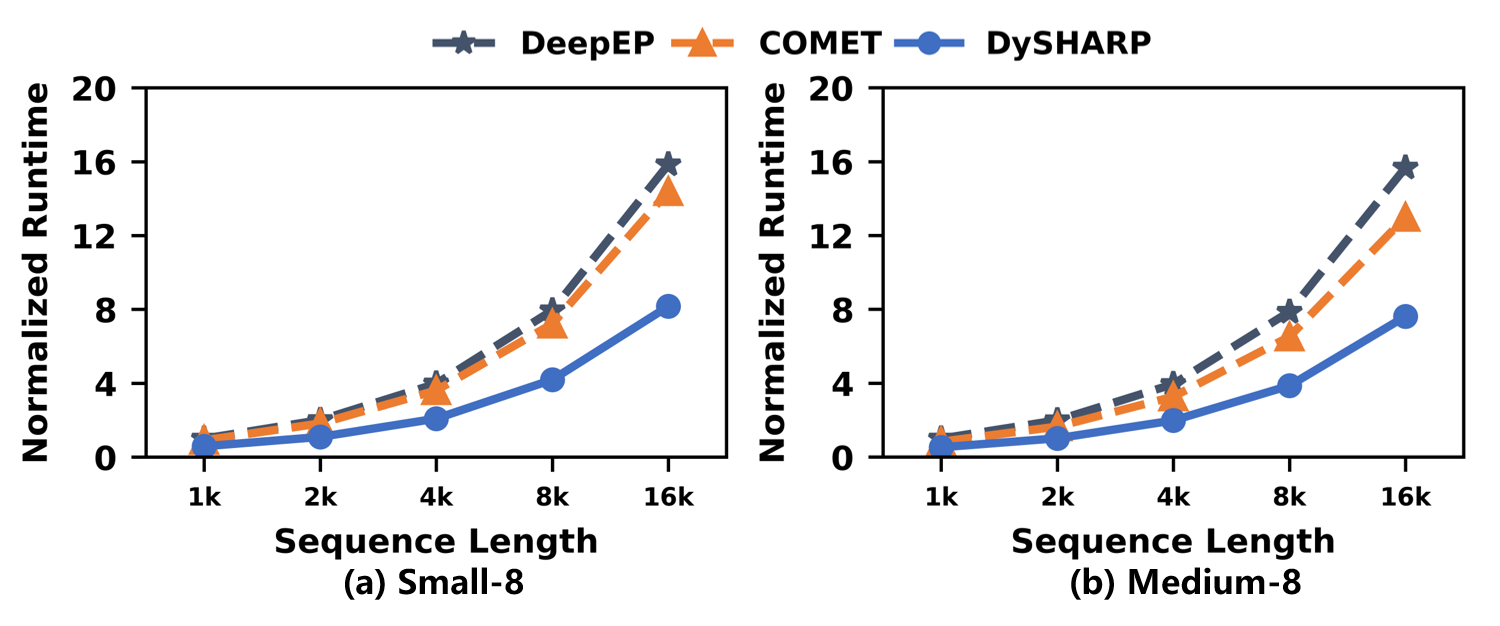}
\vspace{-.2in}
\caption{Performance sensitivity to the sequence length.}
\label{seq}
\vspace{-.17in}
\end{figure}

We further evaluate the performance of DySHARP against baselines across varying sequence lengths. Fig.~\ref{seq} presents the comparison results on MoE layer under sequence lengths of 1024-16384. 
Results demonstrate that DySHARP achieves the shortest execution time regardless of sequence length. As the length increases, the execution times of both DeepEP and COMET rise rapidly, while DySHARP's execution time increases more moderately. This indicates that DySHARP's advantage over baselines becomes more pronounced with longer sequences. 

\subsubsection{Performance of Different Token Distribution}
\label{varystd}

The number of tokens routed to each device is different. Therefore, we evaluate sensitivity to token distribution. 
\reviseB{Following evaluation setup of ByteDance's COMET~\cite{zhang2025comet}, we vary standard deviation of token distribution across experts from 0.01 to 0.05, based on normal distribution $std=0.032$ for a typical training job as introduced in Sec.~\ref{benchmark}.} 
Result demonstrates that DySHARP always achieves remarkable speedups over baselines on MoE layer, regardless of token distribution variations. \looseness=-1

\begin{figure}[!t]
\centering
\includegraphics[width=0.48\textwidth]{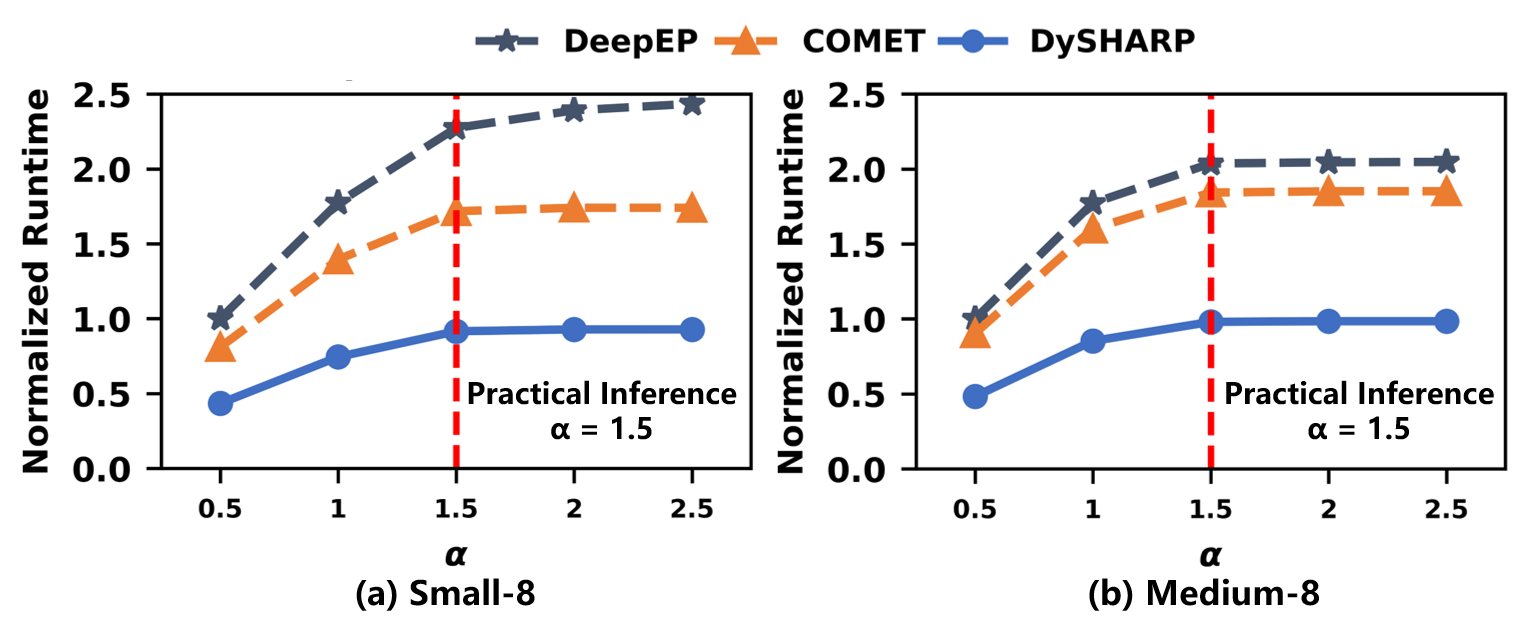}
\vspace{-.2in}
\caption{\reviseB{Performance sensitivity to the token distribution for inference}.}
\label{plaw}
\vspace{-.2in}
\end{figure}

\reviseB{We also evaluate token distribution during inference, different from training~\cite{li2023accelerating}. Our preliminary study reveals a power-law distribution, consistent with recent work~\cite{su2025unveiling} showing that inference token distribution can be modeled as power-law with $\alpha \approx 1.5$. Accordingly, we model inference token distribution as a power-law with $\alpha$ of 0.5-2.5. Fig.~\ref{plaw} shows that, while imbalance prolongs all methods, DySHARP consistently achieves substantial speedup under inference distributions.} \looseness=-1

\begin{figure}[!t]
\centering

\includegraphics[width=0.48\textwidth]{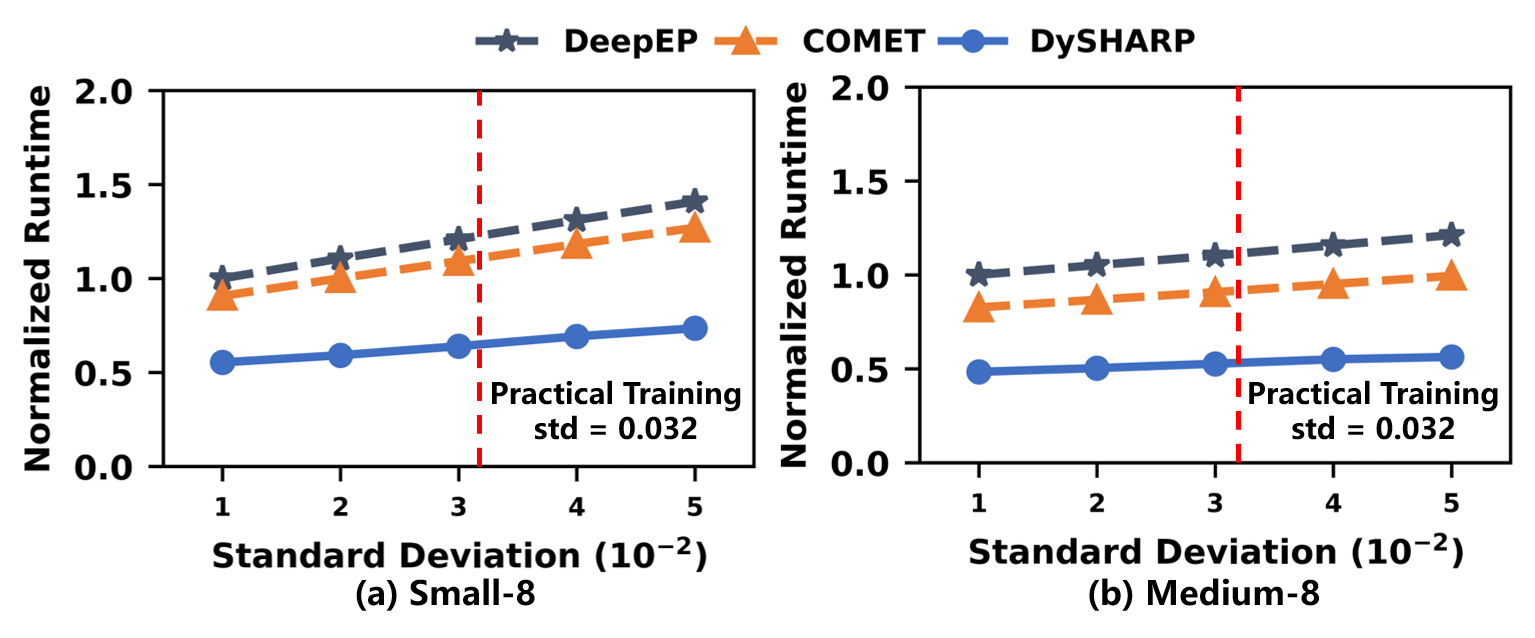}
\vspace{-.2in}
\caption{Performance sensitivity to the token distribution \reviseB{for training}.}
\label{var}
\vspace{-.17in}
\end{figure}

\subsection{Hardware Overhead}
\label{costeval}

We evaluate the hardware overhead of our architectural supports under TSMC 12nm technology~\cite{tsmc}. 
\reviseB{To evaluate hardware overhead, we implement our components in RTL and synthesize them using Synopsys Design Compiler\textsuperscript{\textregistered}~\cite{designcompilier}. SRAM macros are generated with the Memory Compiler of library~\cite{tsmc}. The tables are implemented as 16‑bank dual‑port SRAM (1R1W) to meet DySHARP's concurrent read/write requirements.} 
For the switch, 
\reviseE{by building upon the datapath of existing NVLS, our extension is a lightweight control logic for routing calculation. This enhancement negligibly adds only one cycle to the datapath, without affecting data forwarding.} Area overhead of this logic is less than 0.01$mm^2$. This overhead is less than 0.1\% of NVSwitch die~\cite{hcs2018nvswitch,klenk2020network}. For the GPU, the additional architectural supports require only 0.198$mm^2$, which is about 0.024\% of the H100 GPU die area. The evaluation indicates that our architectural supports are feasible for hardware implementation.

\begin{figure}[!t]
\centering
\includegraphics[width=0.48\textwidth]{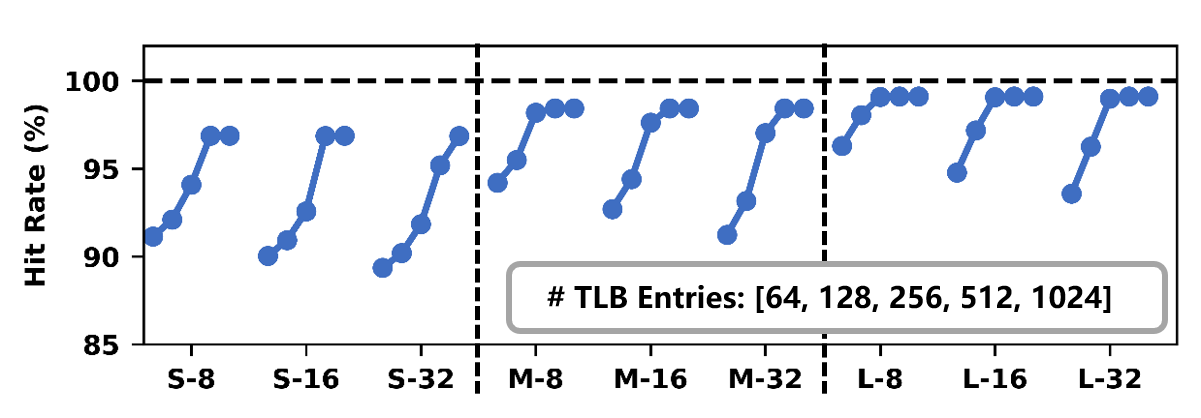}
\vspace{-.17in}
\caption{Design space exploration of AL-TLB.}
\label{tlbdse}
\vspace{-.1in}
\end{figure}

\begin{figure}[!t]
\centering
\includegraphics[width=0.48\textwidth]{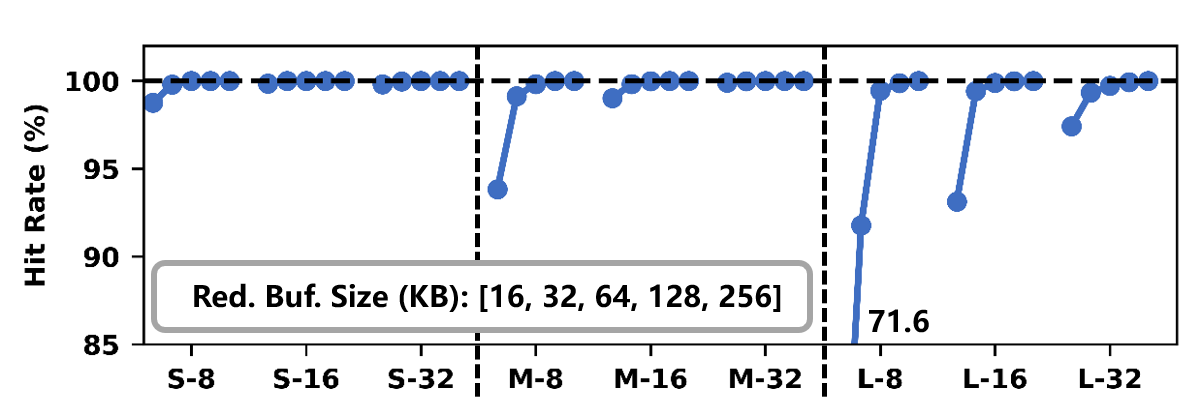}
\vspace{-.17in}
\caption{Design space exploration of reduction buffer.}
\label{redbufdse}
\vspace{-.2in}
\end{figure}

We further explore design space of DySHARP by evaluating hit rates of AL-TLB in Hub and reduction buffer in switch under different sizes. For reduction buffer, a hit means a packet does not trigger eviction. Fig.~\ref{tlbdse} and \ref{redbufdse} show the hit rates of AL-TLB and reduction buffer when varying the sizes. It indicates that 512-entry is a sweet spot for AL-TLB, which can achieve near-ideal hit rates while maintaining small overhead. For the reduction buffer, the sweet spot is 64KB, guaranteeing almost no eviction that increases traffic. 

\section{Discussion}
\label{sec:discussion}

\subsection{\reviseE{Evaluation of End-to-end Inference}}
\label{sec:e2einference}

\begin{figure}[!t]
\centering
\includegraphics[width=0.49\textwidth]{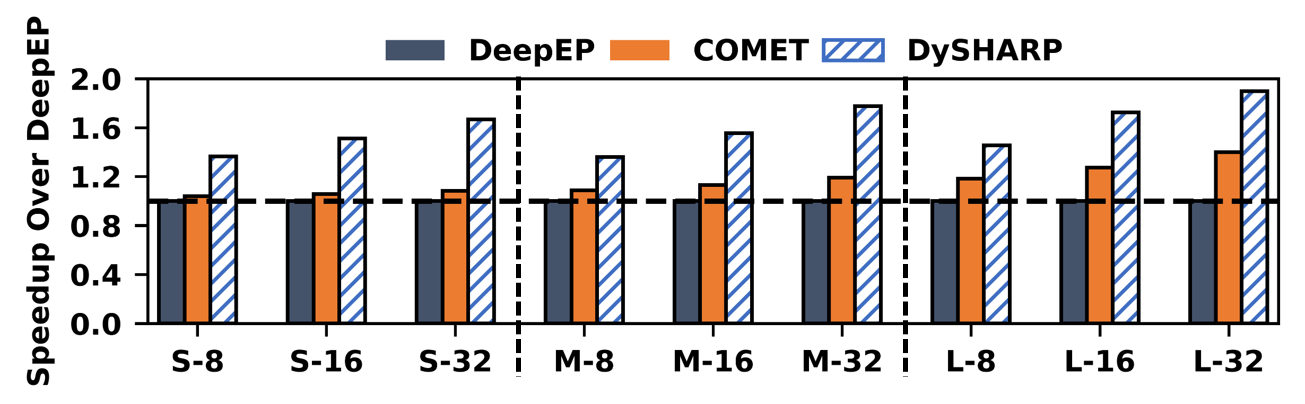}
\vspace{-.27in}
\caption{\reviseE{End-to-end speedup for inference.}}
\label{e2einference}
\vspace{-.15in}
\end{figure}

\reviseE{We further evaluate DySHARP for end-to-end inference, covering both prefill and decode stages. Prefill, like training, is communication-intensive and benefits from DySHARP traffic reduction. Although decode is memory-bound with small batches, its latency sensitivity makes DySHARP's fine-grained synchronization and reduced software control cost impactful. Results in Fig.~\ref{e2einference}, where LLM decodes 512 tokens after prefill, confirm DySHARP's superior inference performance.} \looseness=-1

\subsection{\reviseD{Evaluation on Other Models and Other Platform}}
\label{sec:diverse}

\begin{table}[!h]
      \centering
      \vspace{-.1in}
      \renewcommand{\arraystretch}{1.05}
      \resizebox{0.49\textwidth}{!}{
        \begin{tabular}{|c|c|c|c|c|c|c|}
\hline
\multicolumn{1}{|c|}{Name} & \multicolumn{1}{c|}{\begin{tabular}[c]{@{}c@{}}Hidden\\ Size\end{tabular}} & \multicolumn{1}{c|}{\begin{tabular}[c]{@{}c@{}}MoE Hidden\\ Size\end{tabular}} & \multicolumn{1}{c|}{\begin{tabular}[c]{@{}c@{}}Attention\\ Heads\end{tabular}} & \multicolumn{1}{c|}{\begin{tabular}[c]{@{}c@{}}Sequence\\ Length\end{tabular}} & \multicolumn{1}{c|}{\begin{tabular}[c]{@{}c@{}}Number of\\ Experts\end{tabular}} & \multicolumn{1}{c|}{\begin{tabular}[c]{@{}c@{}}$topk$\\ \end{tabular}} \\ \hline \hline
GPT-OSS-120B & 2880 & 2880 & 64 & 4096 & 64 & 4\\ 
Qwen3-235B & 4096 & 1536 & 128 & 4096 & 128 & 8\\ 
\hline
\end{tabular}
        }
        \label{diversellm}
        \vspace{-.1in}
\end{table}

\begin{figure}[!t]
  \centering
  \begin{minipage}{0.34\textwidth}
    \centering
    \includegraphics[width=\textwidth]{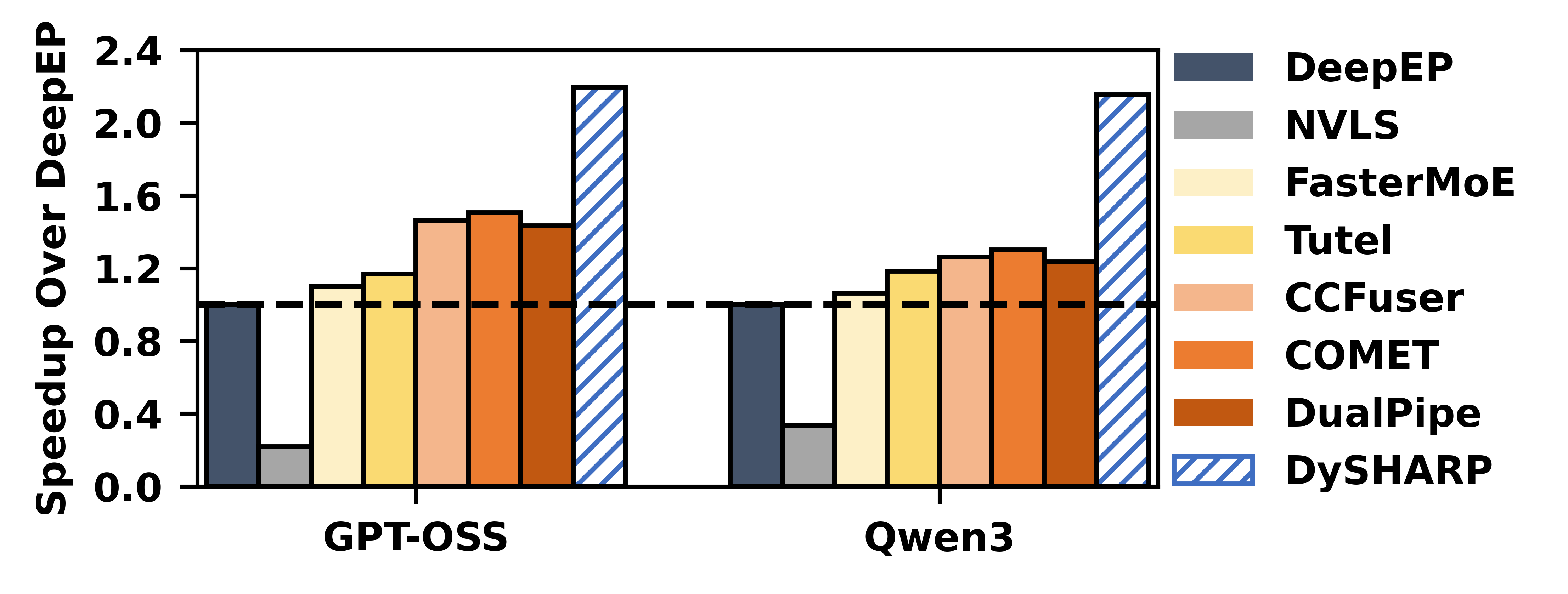}
    \vspace{-.22in}
    \caption{\reviseD{Evaluation of other leading MoE models.}}
    \label{diverse}
  \end{minipage}
  \hfill
  \begin{minipage}{0.14\textwidth}
    \centering
    \hspace{-.25in}
    \includegraphics[width=\textwidth]{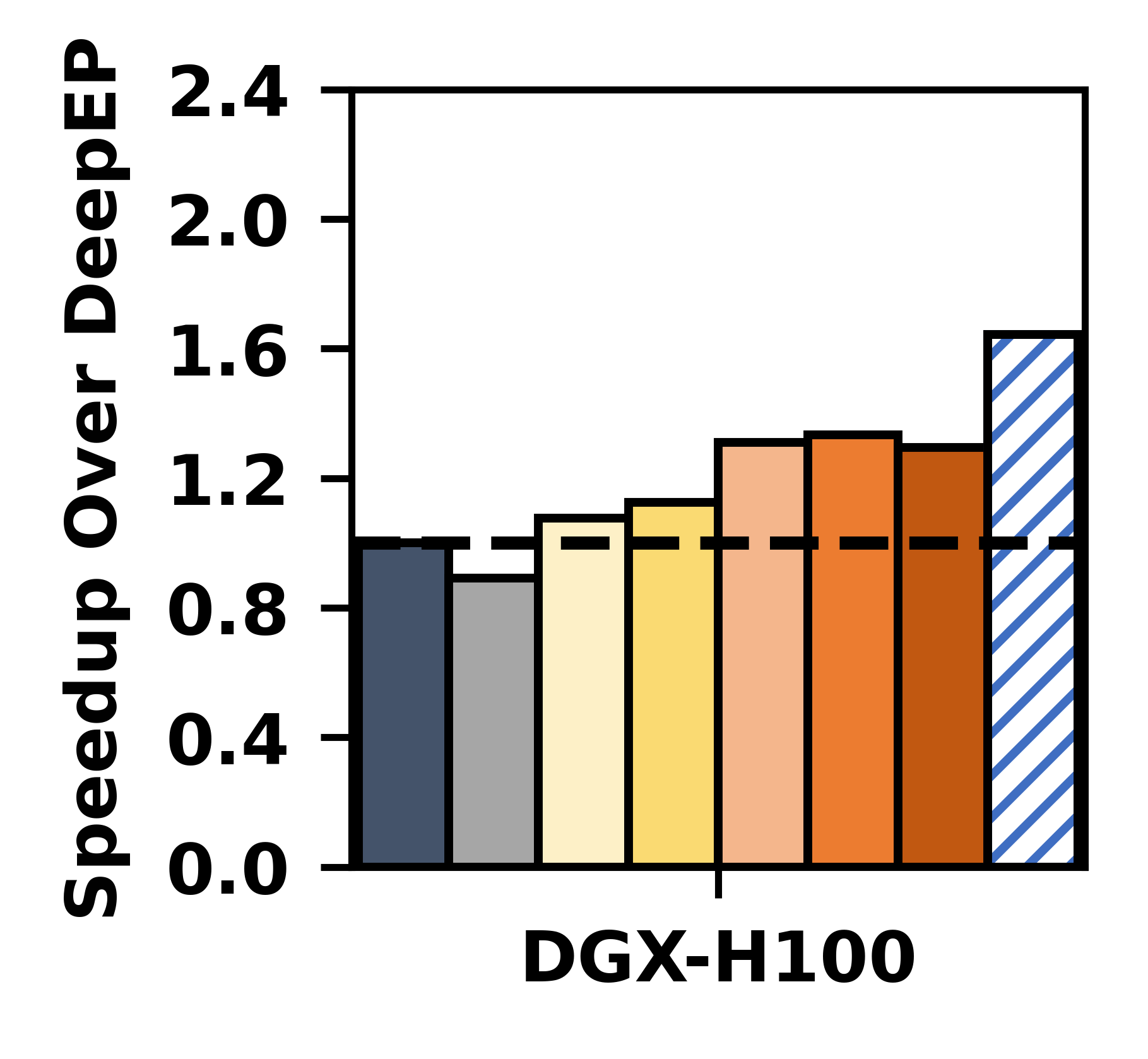}
    \vspace{-.12in}
    \caption{\reviseD{Evaluation on other platform.}}
    \label{platform}
  \end{minipage}
  \vspace{-.2in}
\end{figure}

\reviseD{We evaluate DySHARP for other leading MoE models, including GPT-OSS-120B~\cite{gptoss} and Qwen3-235B~\cite{yang2025qwen3}, as shown in table. Results in Fig.~\ref{diverse} demonstrate DySHARP's superior end-to-end performance on diverse models.}

\reviseD{We evaluate on GH200 NVL32 because single-node systems integrate an increasing number of GPUs~\cite{nvl32,nvl72,rubin}~\cite{nvl32,nvl72,rubin}. DySHARP also applies to regular small nodes like DGX-H100 (8 GPUs). We perform end-to-end evaluation of Large-8 configuration on DGX-H100. Results in Fig.~\ref{platform} confirm DySHARP's advantage on regular small nodes.}

\subsection{\reviseD{Study on Tile Size Choice for Kernel Fusion}}
\label{sec:opttile}

\reviseD{Token-centric kernel fusion adopts the synchronization tile size of 128, the minimum granularity that preserves computation utilization, as it matches the GEMM tile size of 128. A smaller tile would force a suboptimal GEMM tile size and increase synchronization overhead, while a larger tile would coarsen overlap. We validate this on the Small-8 configuration, where smaller models are more sensitive to synchronization granularity. Results in Fig.~\ref{opttile} confirm our choice.} \looseness=-1

\subsection{\reviseD{Extension to Multi-Node System}}
\label{sec:multinode}

InfiniBand (IB)~\cite{ib} networks with Quantum Switch exhibit similar communication redundancy, making multi-node extension feasible. Motivated by the call for a unified network interface~\cite{liu2024deepseek,zhao2025insights}, our extension abstracts the whole cluster as a shared memory system: programmers keep using \texttt{dymultimem.st} and \texttt{dymultimem.ld\_reduce} for cross-node in-switch computing, where NVSwitch and Quantum IB Switch coordinate global routing. Taking multicast (\texttt{dymultimem.st}) as an example, the source GPU duplicates the request: one copy goes to NVSwitch for intra-node delivery, and the other is sent out-of-node, via extended IBGDA~\cite{ibgda} for translation between two communication models, to the IB Switch, which multicasts it to all target nodes. At each remote node, the GPU with the same intra-node ID forwards the request via NVSwitch to destination GPUs. \texttt{dymultimem.ld\_reduce} is similar. Small NVLink packets are aggregated before entering IB to improve performance~\cite{muthukrishnan2023finepack}. Details are omitted for space.

\begin{figure}[!t]
  \centering
  \begin{minipage}{0.19\textwidth}
    \centering
    \includegraphics[width=\textwidth]{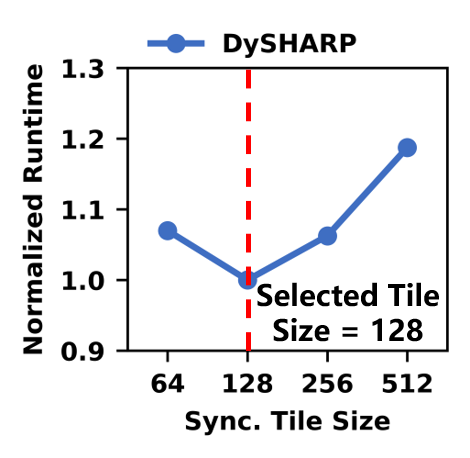}
    \vspace{-.3in}
    \caption{\reviseD{Study on our optimal kernel fusion tile size.}}
    \label{opttile}
  \end{minipage}
  \hfill
  \begin{minipage}{0.28\textwidth}
    \centering
    \vspace{-.05in}
    \includegraphics[width=\textwidth]{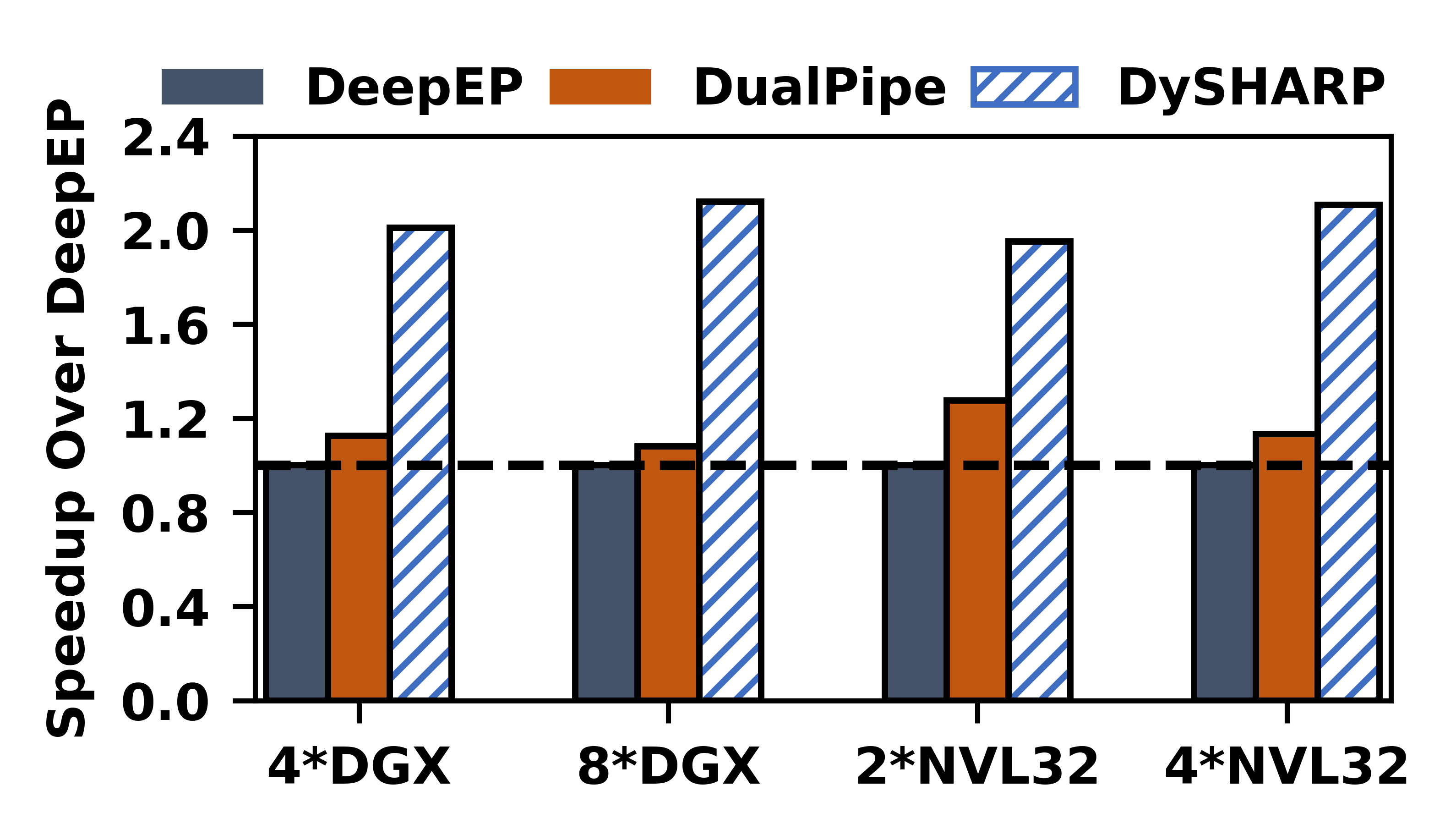}
    \vspace{-.22in}
    \caption{\reviseD{Evaluation of DySHARP's multi-node extension.}}
    \label{multinode}
  \end{minipage}
  \vspace{-.2in}
\end{figure}

\reviseD{We perform preliminary evaluation comparing to DeepEP and DualPipe for end-to-end training (Large-8). We apply expert parallelism across 4/8*DGX-H100 and 2/4*NVL32, and also adopt 16-way pipeline parallelism. Nodes are interconnected with IB. Results in Fig.~\ref{multinode} show the benefit of extended DySHARP over multi-node baselines. While DeepEP reduces inter-node traffic via hierarchical communication \emph{across} intra- and inter-node networks, it cannot eliminate redundancy \emph{within} switch-connected networks (Fig.~\ref{opportunity}) in both intra- and inter-node networks, which DySHARP effectively eliminates.} 

\section{Related Work}
\label{sec:relatedinswitch}

\reviseC{Numerous prior works have proposed leveraging switches to accelerate operations such as AllReduce~\cite{klenk2020network,graham2016scalable,li2019accelerating,gebara2021network,sapio2021scaling,de2021flare,liu2023network,zhang2026towards}, database queries~\cite{lerner2019case,tirmazi2020cheetah}, MapReduce~\cite{chen2021p4com,chen2018programmable}, and DLRM communication~\cite{huang2025traci}. However, most efforts focus on traditional inter-node networks. Only NVLS~\cite{klenk2020network}, CAIS~\cite{zhang2026towards}, and TRACI~\cite{huang2025traci} target inter-chip interconnects with memory semantics. Nevertheless, NVLS and CAIS only support static collectives, and TRACI is tailored to DLRM. Crucially, neither supports the dynamic communication inherent in MoE.} 

An increasing number of works aim to optimize the communication bottleneck in MoE. Some works~\cite{lepikhin2020gshard,nvidiamoe,shen2022se,deepep2025} provide highly optimized libraries. Recent works~\cite{hwang2023tutel,rasley2020deepspeed,nie2022hetumoe,deepep2025} also consider hierarchical communication. 
Beyond optimizing the communication operators in isolation, several works~\cite{he2022fastermoe,hwang2023tutel,zhang2025comet,wang2025harnessing} explore overlapping computation and communication to reduce exposed communication overhead. FasterMoE~\cite{he2022fastermoe} and Tutel~\cite{hwang2023tutel} introduce coarse-grained overlap pipeline. COMET~\cite{zhang2025comet} and CCFuser~\cite{wang2025harnessing} further achieve fine-grained overlap to enhance performance. Some works~\cite{jangda2022breaking,chen2024centauri,pati2024t3,wang2022overlap} target overlapping in dense LLM but are inapplicable to MoE. 
None leverages in-switch computing to reduce communication redundancy in MoE.
\section{Conclusion}

We propose DySHARP, an integral dynamic in-switch computing solution for MoE acceleration. It introduces dynamic multimem addressing and token-centric kernel fusion to achieve fully exploited in-switch computing capabilities. With such a full-stack solution, DySHARP achieves up to 1.79× speedup compared to the SOTA solution.

\section*{Acknowledgement}
We sincerely thank the anonymous ISCA'26 reviewers and shepherd for their valuable suggestions. This work is supported by National Natural Science Foundation of China (NSFC) grant (62502305), Natural Science Foundation of Shanghai (NSFS) grant 25ZR1402275, the Shanghai QiYuan Innovation Foundation QY2025-QN-SJTU-011, Hong Kong Research Grants Council (RGC) CRF-YCRG C6003-24Y, and Shanghai Qi Zhi Institute Innovation Program SQZ202316.


\bibliographystyle{IEEEtranS}
\bibliography{refs}

@inproceedings{klenk2020network,
  title={An in-network architecture for accelerating shared-memory multiprocessor collectives},
  author={Klenk, Benjamin and Jiang, Nan and Thorson, Greg and Dennison, Larry},
  booktitle={2020 ACM/IEEE 47th Annual International Symposium on Computer Architecture (ISCA)},
  pages={996--1009},
  year={2020},
  organization={IEEE}
}

@article{kaplan2020scaling,
  title={Scaling laws for neural language models},
  author={Kaplan, Jared and McCandlish, Sam and Henighan, Tom and Brown, Tom B and Chess, Benjamin and Child, Rewon and Gray, Scott and Radford, Alec and Wu, Jeffrey and Amodei, Dario},
  journal={arXiv preprint arXiv:2001.08361},
  year={2020}
}

@article{cometcode,
  title={Flux: Fine-grained Computation-communication Overlapping GPU Kernel Library.},
  author={ByteDance},
  journal={https://github.com/bytedance/flux},
  year={2025}
}

@article{h100,
  title={NVIDIA H100 Tensor Core GPU.},
  author={NVIDIA},
  journal={https://www.nvidia.com/en-us/data-center/h100},
  year={2022}
}

@article{h200,
  title={NVIDIA H200 Tensor Core GPU.},
  author={NVIDIA},
  journal={https://www.nvidia.com/en-us/data-center/h200},
  year={2023}
}

@article{nvl72,
  title={NVIDIA GB200 NVL72.},
  author={NVIDIA},
  journal={https://www.nvidia.com/en-us/data-center/gb200-nvl72/},
  year={2024}
}

@article{nvl32,
  title={One Giant Superchip for LLMs, Recommenders, and GNNs: Introducing NVIDIA GH200 NVL32.},
  author={NVIDIA},
  journal={https://developer.nvidia.com/blog/one-giant-superchip-for-llms-recommenders-and-gnns-introducing-nvidia-gh200-nvl32},
  year={2023}
}

@article{gb200,
  title={NVIDIA Blackwell Architecture Technical Brief.},
  author={NVIDIA},
  journal={https://resources.nvidia.com/en-us-blackwell-architecture},
  year={2024}
}

@article{rubin,
  title={Inside the NVIDIA Rubin Platform: Six New Chips, One AI Supercomputer.},
  author={NVIDIA},
  journal={https://developer.nvidia.com/blog/inside-the-nvidia-rubin-platform-six-new-chips-one-ai-supercomputer},
  year={2026}
}

@article{dgxh100,
  title={Introduction to NVIDIA DGX H100/H200 Systems.},
  author={NVIDIA},
  journal={https://docs.nvidia.com/dgx/dgxh100-user-guide/introduction-to-dgxh100.html},
  year={2024}
}

@article{gpt5,
  title={Introducing GPT-5.},
  author={OpenAI},
  journal={https://openai.com/index/introducing-gpt-5},
  year={2025}
}

@article{gptoss,
  title={GPT-OSS.},
  author={OpenAI},
  journal={https://github.com/openai/gpt-oss},
  year={2025}
}

@inproceedings{chen2024centauri,
  title={Centauri: Enabling Efficient Scheduling for Communication-Computation Overlap in Large Model Training via Communication Partitioning},
  author={Chen, Chang and Li, Xiuhong and Zhu, Qianchao and Duan, Jiangfei and Sun, Peng and Zhang, Xingcheng and Yang, Chao},
  booktitle={Proceedings of the 29th ACM International Conference on Architectural Support for Programming Languages and Operating Systems, Volume 3},
  pages={178--191},
  year={2024}
}

@inproceedings{jangda2022breaking,
  title={Breaking the computation and communication abstraction barrier in distributed machine learning workloads},
  author={Jangda, Abhinav and Huang, Jun and Liu, Guodong and Sabet, Amir Hossein Nodehi and Maleki, Saeed and Miao, Youshan and Musuvathi, Madanlal and Mytkowicz, Todd and Saarikivi, Olli},
  booktitle={Proceedings of the 27th ACM International Conference on Architectural Support for Programming Languages and Operating Systems},
  pages={402--416},
  year={2022}
}

@inproceedings{pati2024t3,
  title={T3: Transparent Tracking \& Triggering for Fine-grained Overlap of Compute \& Collectives},
  author={Pati, Suchita and Aga, Shaizeen and Islam, Mahzabeen and Jayasena, Nuwan and Sinclair, Matthew D},
  booktitle={Proceedings of the 29th ACM International Conference on Architectural Support for Programming Languages and Operating Systems, Volume 2},
  pages={1146--1164},
  year={2024}
}

@inproceedings{khairy2020accel,
  title={Accel-Sim: An extensible simulation framework for validated GPU modeling},
  author={Khairy, Mahmoud and Shen, Zhesheng and Aamodt, Tor M and Rogers, Timothy G},
  booktitle={2020 ACM/IEEE 47th Annual International Symposium on Computer Architecture (ISCA)},
  pages={473--486},
  year={2020},
  organization={IEEE}
}

@inproceedings{wang2022overlap,
  title={Overlap communication with dependent computation via decomposition in large deep learning models},
  author={Wang, Shibo and Wei, Jinliang and Sabne, Amit and Davis, Andy and Ilbeyi, Berkin and Hechtman, Blake and Chen, Dehao and Murthy, Karthik Srinivasa and Maggioni, Marcello and Zhang, Qiao and Kumar, Sameer and Guo, Tongfei and Xu, Yuanzhong and Zhou, Zongwei},
  booktitle={Proceedings of the 28th ACM International Conference on Architectural Support for Programming Languages and Operating Systems, Volume 1},
  pages={93--106},
  year={2022}
}

@inproceedings{muthukrishnan2023finepack,
  title={Finepack: Transparently improving the efficiency of fine-grained transfers in multi-gpu systems},
  author={Muthukrishnan, Harini and Lustig, Daniel and Villa, Oreste and Wenisch, Thomas and Nellans, David},
  booktitle={2023 IEEE International Symposium on High-Performance Computer Architecture (HPCA)},
  pages={516--529},
  year={2023},
  organization={IEEE}
}

@inproceedings{graham2016scalable,
  title={Scalable hierarchical aggregation protocol (SHArP): A hardware architecture for efficient data reduction},
  author={Graham, Richard L. and Bureddy, Devendar and Lui, Pak and Rosenstock, Hal and Shainer, Gilad and Bloch, Gil and Goldenerg, Dror and Dubman, Mike and Kotchubievsky, Sasha and Koushnir, Vladimir and Levi, Lion and Margolin, Alex and Ronen, Tamir and Shpiner, Alexander and Wertheim, Oded and Zahavi, Eitan},
  booktitle={2016 First International Workshop on Communication Optimizations in HPC (COMHPC)},
  pages={1--10},
  year={2016},
  organization={IEEE}
}

@inproceedings{li2019accelerating,
  title={Accelerating distributed reinforcement learning with in-switch computing},
  author={Li, Youjie and Liu, Iou-Jen and Yuan, Yifan and Chen, Deming and Schwing, Alexander and Huang, Jian},
  booktitle={Proceedings of the 46th International Symposium on Computer Architecture},
  pages={279--291},
  year={2019}
}

@article{gebara2021network,
  title={In-network aggregation for shared machine learning clusters},
  author={Gebara, Nadeen},
  journal={Proceedings of Machine Learning and Systems (MLSys)},
  year={2021}
}

@inproceedings{sapio2021scaling,
  title={Scaling distributed machine learning with $\{$In-Network$\}$ aggregation},
  author={Sapio, Amedeo and Canini, Marco and Ho, Chen-Yu and Nelson, Jacob and Kalnis, Panos and Kim, Changhoon and Krishnamurthy, Arvind and Moshref, Masoud and Ports, Dan and Richt{\'a}rik, Peter},
  booktitle={18th USENIX Symposium on Networked Systems Design and Implementation (NSDI 21)},
  pages={785--808},
  year={2021}
}

@inproceedings{de2021flare,
  title={Flare: Flexible in-network allreduce},
  author={De Sensi, Daniele and Di Girolamo, Salvatore and Ashkboos, Saleh and Li, Shigang and Hoefler, Torsten},
  booktitle={Proceedings of the International Conference for High Performance Computing, Networking, Storage and Analysis},
  pages={1--16},
  year={2021}
}

@inproceedings{liu2023network,
  title={In-network aggregation with transport transparency for distributed training},
  author={Liu, Shuo and Wang, Qiaoling and Zhang, Junyi and Wu, Wenfei and Lin, Qinliang and Liu, Yao and Xu, Meng and Canini, Marco and Cheung, Ray CC and He, Jianfei},
  booktitle={Proceedings of the 28th ACM International Conference on Architectural Support for Programming Languages and Operating Systems, Volume 3},
  pages={376--391},
  year={2023}
}

@inproceedings{hcs2018nvswitch,
  title={NVSwitch and DGX-2},
  author={Ishii, Alex and Foley, Denis and Anderson, Eric and Dally, Bill and Dearth, Glenn and Dennison, Larry and Hummel, Mark and Schafer, John},
  booktitle={Hot Chips},
  year={2018}
}

@inproceedings{rasley2020deepspeed,
  title={Deepspeed: System optimizations enable training deep learning models with over 100 billion parameters},
  author={Rasley, Jeff and Rajbhandari, Samyam and Ruwase, Olatunji and He, Yuxiong},
  booktitle={Proceedings of the 26th ACM SIGKDD international conference on knowledge discovery \& data mining},
  pages={3505--3506},
  year={2020}
}

@inproceedings{jiang2013detailed,
  title={A detailed and flexible cycle-accurate network-on-chip simulator},
  author={Jiang, Nan and Becker, Daniel U and Michelogiannakis, George and Balfour, James and Towles, Brian and Shaw, David E and Kim, John and Dally, William J},
  booktitle={2013 IEEE international symposium on performance analysis of systems and software (ISPASS)},
  pages={86--96},
  year={2013},
  organization={IEEE}
}

@article{chen2021p4com,
  title={P4com: In-network computation with programmable switches},
  author={Chen, Ge and Zeng, Gaoxiong and Chen, Li},
  journal={arXiv preprint arXiv:2107.13694},
  year={2021}
}

@article{chen2018programmable,
  title={Programmable switch as a parallel computing device},
  author={Chen, Li and Chen, Ge and Lingys, Justinas and Chen, Kai},
  journal={arXiv preprint arXiv:1803.01491},
  year={2018}
}

@inproceedings{lerner2019case,
  title={The Case for Network Accelerated Query Processing.},
  author={Lerner, Alberto and Hussein, Rana and Cudre-Mauroux, Philippe and eXascale Infolab, U},
  booktitle={CIDR},
  year={2019}
}

@inproceedings{tirmazi2020cheetah,
  title={Cheetah: Accelerating database queries with switch pruning},
  author={Tirmazi, Muhammad and Ben Basat, Ran and Gao, Jiaqi and Yu, Minlan},
  booktitle={Proceedings of the 2020 ACM SIGMOD International Conference on Management of Data},
  pages={2407--2422},
  year={2020}
}

@inproceedings{huang2025traci,
  title={TRACI: Network Acceleration of Input-Dynamic Communication for Large-Scale Deep Learning Recommendation Model},
  author={Huang, Guyue and Li, Hao and Qin, Le and Huang, Jiayi and Kang, Yangwook and Ding, Yufei and Xie, Yuan},
  booktitle={Proceedings of the 52nd Annual International Symposium on Computer Architecture},
  pages={1880--1893},
  year={2025}
}

@article{lepikhin2020gshard,
  title={Gshard: Scaling giant models with conditional computation and automatic sharding},
  author={Lepikhin, Dmitry and Lee, HyoukJoong and Xu, Yuanzhong and Chen, Dehao and Firat, Orhan and Huang, Yanping and Krikun, Maxim and Shazeer, Noam and Chen, Zhifeng},
  journal={arXiv preprint arXiv:2006.16668},
  year={2020}
}

@article{nvidiamoe,
  title={Doubling all2all performance with nvidia collective communication library 2.12},
  author={NVIDIA},
  journal={ https://developer.nvidia.com/blog/doubling-all2all-performance-with/nvidia-collective-communication/library-2-12/},
  year={2022}
}

@article{shen2022se,
  title={Se-moe: A scalable and efficient mixture-of-experts distributed training and inference system},
  author={Shen, Liang and Wu, Zhihua and Gong, WeiBao and Hao, Hongxiang and Bai, Yangfan and Wu, HuaChao and Wu, Xinxuan and Bian, Jiang and Xiong, Haoyi and Yu, Dianhai and Ma, Yanjun},
  journal={arXiv e-prints},
  pages={arXiv--2205},
  year={2022}
}

@article{hwang2023tutel,
  title={Tutel: Adaptive mixture-of-experts at scale},
  author={Hwang, Changho and Cui, Wei and Xiong, Yifan and Yang, Ziyue and Liu, Ze and Hu, Han and Wang, Zilong and Salas, Rafael and Jose, Jithin and Ram, Prabhat and Chau, HoYuen and Cheng, Peng and Yang, Fan and Yang, Mao and Xiong, Yongqiang},
  journal={Proceedings of Machine Learning and Systems},
  volume={5},
  pages={269--287},
  year={2023}
}

@inproceedings{rajbhandari2022deepspeed,
  title={Deepspeed-moe: Advancing mixture-of-experts inference and training to power next-generation ai scale},
  author={Rajbhandari, Samyam and Li, Conglong and Yao, Zhewei and Zhang, Minjia and Aminabadi, Reza Yazdani and Awan, Ammar Ahmad and Rasley, Jeff and He, Yuxiong},
  booktitle={International conference on machine learning},
  pages={18332--18346},
  year={2022},
  organization={PMLR}
}

@article{nie2022hetumoe,
  title={HetuMoE: An efficient trillion-scale mixture-of-expert distributed training system},
  author={Nie, Xiaonan and Zhao, Pinxue and Miao, Xupeng and Zhao, Tong and Cui, Bin},
  journal={arXiv preprint arXiv:2203.14685},
  year={2022}
}

@misc{deepep2025,
      title={DeepEP: an efficient expert-parallel communication library},
      author={Chenggang Zhao and Shangyan Zhou and Liyue Zhang and Chengqi Deng and Zhean Xu and Yuxuan Liu and Kuai Yu and Jiashi Li and Liang Zhao},
      year={2025},
      publisher = {GitHub},
      howpublished = {\url{https://github.com/deepseek-ai/DeepEP}},
}

@inproceedings{he2022fastermoe,
  title={Fastermoe: modeling and optimizing training of large-scale dynamic pre-trained models},
  author={He, Jiaao and Zhai, Jidong and Antunes, Tiago and Wang, Haojie and Luo, Fuwen and Shi, Shangfeng and Li, Qin},
  booktitle={Proceedings of the 27th ACM SIGPLAN Symposium on Principles and Practice of Parallel Programming},
  pages={120--134},
  year={2022}
}

@article{zhang2025comet,
  title={Comet: Fine-grained computation-communication overlapping for mixture-of-experts},
  author={Zhang, Shulai and Zheng, Ningxin and Lin, Haibin and Jiang, Ziheng and Bao, Wenlei and Jiang, Chengquan and Hou, Qi and Cui, Weihao and Zheng, Size and Chang, Li-Wen and Chen, Quan and Liu, Xin},
  journal={arXiv preprint arXiv:2502.19811},
  year={2025}
}

@inproceedings{wang2025harnessing,
  title={Harnessing inter-gpu shared memory for seamless moe communication-computation fusion},
  author={Wang, Hulin and Xia, Yaqi and Yang, Donglin and Zhou, Xiaobo and Cheng, Dazhao},
  booktitle={Proceedings of the 30th ACM SIGPLAN Annual Symposium on Principles and Practice of Parallel Programming},
  pages={170--182},
  year={2025}
}

@article{vaswani2017attention,
  title={Attention is all you need},
  author={Vaswani, Ashish and Shazeer, Noam and Parmar, Niki and Uszkoreit, Jakob and Jones, Llion and Gomez, Aidan N and Kaiser, {\L}ukasz and Polosukhin, Illia},
  journal={Advances in neural information processing systems},
  volume={30},
  year={2017}
}

@article{liu2024deepseek,
  title={Deepseek-v3 technical report},
  author={DeepSeek-AI and Aixin Liu and Bei Feng and Bing Xue and Bingxuan Wang and Bochao Wu and Chengda Lu and Chenggang Zhao and Chengqi Deng and Chenyu Zhang and Chong Ruan and Damai Dai and Daya Guo and Dejian Yang and Deli Chen and Dongjie Ji and Erhang Li and Fangyun Lin and Fucong Dai and Fuli Luo and Guangbo Hao and Guanting Chen and Guowei Li and H. Zhang and Han Bao and Hanwei Xu and Haocheng Wang and Haowei Zhang and Honghui Ding and Huajian Xin and Huazuo Gao and Hui Li and Hui Qu and J. L. Cai and Jian Liang and Jianzhong Guo and Jiaqi Ni and Jiashi Li and Jiawei Wang and Jin Chen and Jingchang Chen and Jingyang Yuan and Junjie Qiu and Junlong Li and Junxiao Song and Kai Dong and Kai Hu and Kaige Gao and Kang Guan and Kexin Huang and Kuai Yu and Lean Wang and Lecong Zhang and Lei Xu and Leyi Xia and Liang Zhao and Litong Wang and Liyue Zhang and Meng Li and Miaojun Wang and Mingchuan Zhang and Minghua Zhang and Minghui Tang and Mingming Li and Ning Tian and Panpan Huang and Peiyi Wang and Peng Zhang and Qiancheng Wang and Qihao Zhu and Qinyu Chen and Qiushi Du and R. J. Chen and R. L. Jin and Ruiqi Ge and Ruisong Zhang and Ruizhe Pan and Runji Wang and Runxin Xu and Ruoyu Zhang and Ruyi Chen and S. S. Li and Shanghao Lu and Shangyan Zhou and Shanhuang Chen and Shaoqing Wu and Shengfeng Ye and Shengfeng Ye and Shirong Ma and Shiyu Wang and Shuang Zhou and Shuiping Yu and Shunfeng Zhou and Shuting Pan and T. Wang and Tao Yun and Tian Pei and Tianyu Sun and W. L. Xiao and Wangding Zeng and Wanjia Zhao and Wei An and Wen Liu and Wenfeng Liang and Wenjun Gao and Wenqin Yu and Wentao Zhang and X. Q. Li and Xiangyue Jin and Xianzu Wang and Xiao Bi and Xiaodong Liu and Xiaohan Wang and Xiaojin Shen and Xiaokang Chen and Xiaokang Zhang and Xiaosha Chen and Xiaotao Nie and Xiaowen Sun and Xiaoxiang Wang and Xin Cheng and Xin Liu and Xin Xie and Xingchao Liu and Xingkai Yu and Xinnan Song and Xinxia Shan and Xinyi Zhou and Xinyu Yang and Xinyuan Li and Xuecheng Su and Xuheng Lin and Y. K. Li and Y. Q. Wang and Y. X. Wei and Y. X. Zhu and Yang Zhang and Yanhong Xu and Yanhong Xu and Yanping Huang and Yao Li and Yao Zhao and Yaofeng Sun and Yaohui Li and Yaohui Wang and Yi Yu and Yi Zheng and Yichao Zhang and Yifan Shi and Yiliang Xiong and Ying He and Ying Tang and Yishi Piao and Yisong Wang and Yixuan Tan and Yiyang Ma and Yiyuan Liu and Yongqiang Guo and Yu Wu and Yuan Ou and Yuchen Zhu and Yuduan Wang and Yue Gong and Yuheng Zou and Yujia He and Yukun Zha and Yunfan Xiong and Yunxian Ma and Yuting Yan and Yuxiang Luo and Yuxiang You and Yuxuan Liu and Yuyang Zhou and Z. F. Wu and Z. Z. Ren and Zehui Ren and Zhangli Sha and Zhe Fu and Zhean Xu and Zhen Huang and Zhen Zhang and Zhenda Xie and Zhengyan Zhang and Zhewen Hao and Zhibin Gou and Zhicheng Ma and Zhigang Yan and Zhihong Shao and Zhipeng Xu and Zhiyu Wu and Zhongyu Zhang and Zhuoshu Li and Zihui Gu and Zijia Zhu and Zijun Liu and Zilin Li and Ziwei Xie and Ziyang Song and Ziyi Gao and Zizheng Pan},
  journal={arXiv preprint arXiv:2412.19437},
  year={2024}
}

@article{yang2025qwen3,
  title={Qwen3 technical report},
  author={An Yang and Anfeng Li and Baosong Yang and Beichen Zhang and Binyuan Hui and Bo Zheng and Bowen Yu and Chang Gao and Chengen Huang and Chenxu Lv and Chujie Zheng and Dayiheng Liu and Fan Zhou and Fei Huang and Feng Hu and Hao Ge and Haoran Wei and Huan Lin and Jialong Tang and Jian Yang and Jianhong Tu and Jianwei Zhang and Jianxin Yang and Jiaxi Yang and Jing Zhou and Jingren Zhou and Junyang Lin and Kai Dang and Keqin Bao and Kexin Yang and Le Yu and Lianghao Deng and Mei Li and Mingfeng Xue and Mingze Li and Pei Zhang and Peng Wang and Qin Zhu and Rui Men and Ruize Gao and Shixuan Liu and Shuang Luo and Tianhao Li and Tianyi Tang and Wenbiao Yin and Xingzhang Ren and Xinyu Wang and Xinyu Zhang and Xuancheng Ren and Yang Fan and Yang Su and Yichang Zhang and Yinger Zhang and Yu Wan and Yuqiong Liu and Zekun Wang and Zeyu Cui and Zhenru Zhang and Zhipeng Zhou and Zihan Qiu},
  journal={arXiv preprint arXiv:2505.09388},
  year={2025}
}

@article{tang2025pangu,
  title={Pangu ultra moe: How to train your big moe on ascend npus},
  author={Yehui Tang and Yichun Yin and Yaoyuan Wang and Hang Zhou and Yu Pan and Wei Guo and Ziyang Zhang and Miao Rang and Fangcheng Liu and Naifu Zhang and Binghan Li and Yonghan Dong and Xiaojun Meng and Yasheng Wang and Dong Li and Yin Li and Dandan Tu and Can Chen and Youliang Yan and Fisher Yu and Ruiming Tang and Yunhe Wang and Botian Huang and Bo Wang and Boxiao Liu and Changzheng Zhang and Da Kuang and Fei Liu and Gang Huang and Jiansheng Wei and Jiarui Qin and Jie Ran and Jinpeng Li and Jun Zhao and Liang Dai and Lin Li and Liqun Deng and Peifeng Qin and Pengyuan Zeng and Qiang Gu and Shaohua Tang and Shengjun Cheng and Tao Gao and Tao Yu and Tianshu Li and Tianyu Bi and Wei He and Weikai Mao and Wenyong Huang and Wulong Liu and Xiabing Li and Xianzhi Yu and Xueyu Wu and Xu He and Yangkai Du and Yan Xu and Ye Tian and Yimeng Wu and Yongbing Huang and Yong Tian and Yong Zhu and Yue Li and Yufei Wang and Yuhang Gai and Yujun Li and Yu Luo and Yunsheng Ni and Yusen Sun and Zelin Chen and Zhe Liu and Zhicheng Liu and Zhipeng Tu and Zilin Ding and Zongyuan Zhan},
  journal={arXiv preprint arXiv:2505.04519},
  year={2025}
}

@inproceedings{liu2021swin,
  title={Swin transformer: Hierarchical vision transformer using shifted windows},
  author={Liu, Ze and Lin, Yutong and Cao, Yue and Hu, Han and Wei, Yixuan and Zhang, Zheng and Lin, Stephen and Guo, Baining},
  booktitle={Proceedings of the IEEE/CVF international conference on computer vision},
  pages={10012--10022},
  year={2021}
}

@article{dosovitskiy2020image,
  title={An image is worth 16x16 words: Transformers for image recognition at scale},
  author={Alexey Dosovitskiy and Lucas Beyer and Alexander Kolesnikov and Dirk Weissenborn and Xiaohua Zhai and Thomas Unterthiner and Mostafa Dehghani and Matthias Minderer and Georg Heigold and Sylvain Gelly and Jakob Uszkoreit and Neil Houlsby},
  journal={arXiv preprint arXiv:2010.11929},
  year={2020}
}

@inproceedings{ishii2022nvlink,
  title={The nvlink-network switch: Nvidia’s switch chip for high communication-bandwidth superpods},
  author={Ishii, Alexander and Wells, Ryan},
  booktitle={2022 IEEE Hot Chips 34 Symposium (HCS)},
  pages={1--23},
  year={2022},
  organization={IEEE Computer Society}
}

@inproceedings{zhao2025insights,
  title={Insights into DeepSeek-V3: Scaling Challenges and Reflections on Hardware for AI Architectures},
  author={Zhao, Chenggang and Deng, Chengqi and Ruan, Chong and Dai, Damai and Gao, Huazuo and Li, Jiashi and Zhang, Liyue and Huang, Panpan and Zhou, Shangyan and Ma, Shirong and Liang, Wenfeng and He, Ying and Wang, Yuqing and Liu, Yuxuan and Wei, Y.X.},
  booktitle={2025 ACM/IEEE 52nd Annual International Symposium on Computer Architecture (ISCA)},
  pages = {1731–1745},
  year={2025},
  organization={ACM}
}

@article{llama4,
  title={The Llama 4 herd: The beginning of a new era of natively multimodal AI innovation},
  author={Meta},
  journal={https://ai.meta.com/blog/llama-4-multimodal-intelligence},
  year={2025}
}

@inproceedings{li2023accelerating,
  title={Accelerating distributed $\{$MoE$\}$ training and inference with lina},
  author={Li, Jiamin and Jiang, Yimin and Zhu, Yibo and Wang, Cong and Xu, Hong},
  booktitle={2023 USENIX Annual Technical Conference (USENIX ATC 23)},
  pages={945--959},
  year={2023}
}

@article{su2025unveiling,
  title={Unveiling super experts in mixture-of-experts large language models},
  author={Su, Zunhai and Li, Qingyuan and Zhang, Hao and Ye, Weihao and Xue, Qibo and Qian, YuLei and Xie, Yuchen and Wong, Ngai and Yuan, Kehong},
  journal={arXiv preprint arXiv:2507.23279},
  year={2025}
}

@article{designcompilier,
  title={Design Compiler® RTL Synthesis.},
  author={Synopsys},
  journal={https://www.synopsys.com/implementation-and-signoff/rtl-synthesis-test/design-compiler-nxt.html},
  year={2021}
}

@article{tsmc,
  title={TSMC 16nm and 12nm process technologies.},
  author={TSMC},
  journal={https://www.tsmc.com/english/dedicatedFoundry/technology/logic/l\_16\_12nm},
  year={2017}
}

@article{ib,
  title={The NVIDIA Quantum InfiniBand Platform.},
  author={NVIDIA},
  journal={https://www.nvidia.com/en-us/networking/products/infiniband},
  year={2025}
}

@article{ibgda,
  title={Improving Network Performance of HPC Systems Using NVIDIA Magnum IO NVSHMEM and GPUDirect Async.},
  author={NVIDIA},
  journal={https://developer.nvidia.com/blog/improving-network-performance-of-hpc-systems-using-nvidia-magnum-io-nvshmem-and-gpudirect-async},
  year={2025}
}

@inproceedings{ma2020rammer,
  title={Rammer: Enabling holistic deep learning compiler optimizations with $\{$rTasks$\}$},
  author={Ma, Lingxiao and Xie, Zhiqiang and Yang, Zhi and Xue, Jilong and Miao, Youshan and Cui, Wei and Hu, Wenxiang and Yang, Fan and Zhang, Lintao and Zhou, Lidong},
  booktitle={14th USENIX Symposium on Operating Systems Design and Implementation (OSDI 20)},
  pages={881--897},
  year={2020}
}

@inproceedings{zhang2026towards,
  title={Towards Compute-Aware In-Switch Computing for LLMs Tensor-Parallelism on Multi-GPU Systems},
  author={Zhang, Chen and Zhang, Qijun and Zhou, Zhuoshan and Diao, Yijia and Wang, Haibo and Zhou, Zhe and Tu, Zhipeng and Li, Zhiyao and Sun, Guangyu and Song, Zhuoran and others},
  booktitle={2026 IEEE International Symposium on High Performance Computer Architecture (HPCA)},
  pages={1--15},
  year={2026},
  organization={IEEE}
}

\end{document}